\documentclass[twocolumn,twocolappendix]{aastex701}
\usepackage{amsmath}
\usepackage{ulem}
\begin{document}

\title{Systematic Comparison between Constrained Transport and Mixed Divergence Cleaning Methods for Astrophysical Magnetohydrodynamic Simulations}

\author[orcid=0000-0001-8105-8113,gname=Kengo,sname=Tomida]{Kengo Tomida}
\affiliation{Astronomical Institute, Tohoku University, Sendai, Miyagi 980-8578, Japan}
\email[show]{tomida@astr.tohoku.ac.jp}  

\author[orcid=0009-0001-8312-8506, gname='Kenji Eric',sname=Sadanari]{Kenji Eric Sadanari} 
\affiliation{\'{E}cole Normale Sup\'erieure de Lyon, Centre de Recherche Astrophysique de Lyon, UMR5574, 69007 Lyon, France}
\email{kenji.sadanari@ens-lyon.fr}

\author[orcid=0000-0003-3882-3945,gname=Shinsuke,sname=Takasao]{Shinsuke Takasao}
\affiliation{Humanities and Sciences/Museum Careers, Musashino Art University, Kodaira, Tokyo 187-8505, Japan}
\email{stakasao@musabi.ac.jp}

\author[orcid=0000-0002-2707-7548,gname=Kazunari,sname=Iwasaki]{Kazunari Iwasaki}
\affiliation{Center for Computational Astrophysics, National Astronomical Observatory of Japan, Mitaka, Tokyo 181-8588, Japan}
\email{kazunari.iwasaki@nao.ac.jp}

\begin{abstract}
Magnetohydrodynamic (MHD) simulations are indispensable research infrastructure in astrophysics today. In order to satisfy the solenoidal constraint of the MHD equations on discretized grids, modern simulation codes often employ either constrained transport (CT) with a staggered grid or divergence cleaning using an additional variable. We compare CT and Dedner's mixed divergence cleaning schemes systematically, and find that the divergence cleaning scheme can produce substantial artifacts in certain situations. Through numerical experiments including both idealized tests and practical applications, we show that the original implementation of Dedner's scheme becomes inaccurate when magnetic fields are strongly localized or when the timestep suddenly changes. We find that some previous results, such as the extremely rapid growth of magnetic fields during star formation in the early Universe, may be affected by the spurious behavior of the divergence cleaning scheme. We propose a few modifications to improve the robustness of the divergence cleaning method. Nevertheless, we find that the CT scheme is more accurate and reliable in many situations.
\end{abstract}


\section{Introduction} 
Magnetohydrodynamic (MHD) simulations are used to study various astrophysical applications such as black hole accretion disks, galactic interstellar media, star formation, protoplanetary disks, and so on. A variety of MHD simulation codes are now publicly available, including grid-based codes such as Athena++ \citep{athenapp}, AstroBEAR \citep{astrobear}, CANS+ \citep{cans}, Enzo \citep{enzo}, FLASH \citep{flash}, MPI-AMRVAC \citep{amrvac}, PLUTO \citep{pluto}, RAMSES \citep{ramsesmhd} and RIEMANN \citep{balsara01,balsara04b}, and particle-based (including smoothed-particle hydrodynamics (SPH), moving-mesh and meshless methods) codes such as AREPO \citep{arepo,arepoct}, Gadget \citep{gadgetmhd}, GIZMO \citep{gizmo} and Phantom \citep{phantom} \citep[see also][]{iwasaki13}.

In the MHD equations, the solenoidal constraint ($\nabla\cdot\mathbf{B}=0$) is automatically satisfied if the initial condition satisfies it. 
However, this constraint is not trivially satisfied in the discretized MHD equations. If $\nabla\cdot\mathbf{B}$ becomes unphysically large, it can cause spurious acceleration along magnetic fields and produce numerical artifacts, sometimes leading simulations to crash\footnote{If the simulation crashes, at least we can tell something has gone wrong. An even worse case is when the simulation continues running without crashing while producing unphysical results that are difficult to detect.}.  For stable MHD simulations, some techniques are necessary to maintain $\nabla\cdot\mathbf{B}$ sufficiently small and suppress artificial forces associated with $\nabla\cdot\mathbf{B}$. There are roughly three kinds of such techniques: projection, divergence cleaning, and constrained transport (CT).

In the projection method (\citet{brackbill80}, see also \citet{balsara04}), we use the ``potential" $\phi$ to cancel out the divergence as:
\begin{eqnarray}
\nabla^2\phi&=&\nabla\cdot\mathbf{B},\\
\mathbf{B}'&=&\mathbf{B}-\nabla\phi.
\end{eqnarray}
This requires solving the Poisson equation. While it can be solved using established methods such as Fast Fourier Transform (on uniform grids) \citep{balsara04} and Multigrid (suitable for mesh refinement) \citep{teunissen19}, it is computationally expensive because it requires a global solution. Therefore, this method is not commonly supported by public MHD codes.

Most modern MHD simulation codes support either the divergence cleaning or CT scheme. The divergence cleaning methods are popular because they are easy to implement, compatible with higher-order methods and mesh refinement, computationally efficient, and (sometimes excessively) robust. There are a few different methods in this category including Powell's 8-wave method \citep[][see Appendix~\ref{sec:powell}]{powell94,powell}, and among them, the mixed-cleaning using the generalized Lagrange multiplier (GLM) proposed by \citet{dedner} is arguably the most popular. In this scheme, the additional variable $\psi$ associated with $\nabla\cdot\mathbf{B}$ is introduced, and the divergence error is transported while being damped. This method is adopted by many public simulation codes, and is the primary choice for particle-based codes because the CT scheme is not trivial for irregular particle distributions \citep[but see e.g.][]{arepoct}.

The CT scheme \citep{ct}, on the other hand, utilizes staggered grids and defines the magnetic fields on the surfaces of each cell. The magnetic fields are updated with the electromotive forces (EMFs) calculated on the cell edges using the Stokes' theorem. While this scheme is often less diffusive and more accurate compared to the divergence cleaning, it tends to be more expensive per timestep than the divergence cleaning as it requires additional operations for updating the magnetic fields. The staggered grid is more complicated to implement, and its implementation with mesh refinement requires considerable care. Moreover, it can be less robust in the low plasma-beta (strong magnetization) limit, because the scheme cannot guarantee the gas pressure to be positive because the magnetic fields and the gas energy reside at different positions. Usually the cell-centered fields are calculated by averaging the face-centered fields with some weights, while divergence-free reconstruction \citep{balsara04b} can improve the stability. It is also known to be unstable particularly when numerical reconnection occurs in a strongly magnetized region \citep{hawley95}. Moreover, because it is not a scheme to enforce $\nabla\cdot\mathbf{B}=0$ but to conserve $\nabla\cdot\mathbf{B}$, the magnetic fields must be carefully initialized. To satisfy $\nabla\cdot\mathbf{B}=0$ in the initial condition, one must initialize the fields using circular integration of the vector potential on cell edges. This makes it difficult to start simulations with non-trivial magnetic field configurations with the CT scheme. Nevertheless, this scheme has a fundamental advantage; it is the natural discretization of the integral form of the induction equation, and it can conserve $\nabla\cdot\mathbf{B}$ to machine precision. 

Both schemes have been extensively tested and demonstrated to work well in simple idealized test problems \citep{toth00,balsara04,vide13,zhang16} as well as some practical astrophysical applications such as star formation, interstellar media, solar atmosphere, accretion disks, and so on. In general, the divergence cleaning tends to be more diffusive compared to the CT scheme \citep[e.g.][]{vide13}, but both schemes yield reasonably consistent results. However, it is not clear whether the divergence cleaning schemes remain accurate in numerically difficult situations. For example, \citet{hirano22} \citep[see also][]{machida25} reported that magnetic fields are exponentially amplified from tiny seed fields by orders of magnitude within a few orbital time scales in the context of primordial star formation. In contrast, \citet{sadanari21,sadanari23,sadanari24} did not find such a rapid amplification of magnetic fields in a similar context, although the details of the simulation setups are different. The fact that all of these works used the mixed divergence cleaning scheme motivates us to investigate the behavior of the divergence cleaning in challenging situations in comparison with the CT scheme. 

In this paper, we compare the mixed divergence cleaning and CT schemes systematically. For this purpose, we implement the mixed divergence cleaning scheme on the Athena++ code in addition to the existing CT solver. This enables a fair and consistent comparison between the two schemes. Using the code, we identify certain situations where the divergence cleaning scheme gives inaccurate solutions, and also propose some improvements to it. The rest of the paper is organized as follows. We review the divergence cleaning and CT schemes in Section 2. We also point out possible issues with the original implementation of the mixed divergence cleaning and propose a few improvements. In Section 3, we present the results of the comparison to demonstrate certain situations where the divergence cleaning method fails. Section 4 is devoted to discussions, and Section 5 summarizes our findings.

\section{Methods} \label{sec:methods}
\subsection{Basic Equations}
Throughout this paper, we use the ideal MHD equations with the magnetic permeability $\mu=1$.
\begin{eqnarray}
\frac{\partial \rho}{\partial t} + \mathbf{\nabla\cdot} (\rho\mathbf{ v}) &=& 0,\label{eq:eoc}\\
\frac{\partial \rho \mathbf{v}}{\partial t} +\mathbf{\nabla\cdot} \left(\rho\mathbf{vv} - \mathbf{BB} + P^*{\mathbb I}\right)  &=& 0,\label{eq:eom}\\
\frac{\partial \mathbf{B}}{\partial t} - \mathbf{\nabla} \times \left(\mathbf{ v} \times \mathbf{B}\right) &=&0,\label{eq:ind}\\
\nabla\cdot\mathbf{B}&=&0,\label{eq:sol}\\
\frac{\partial E}{\partial t} +\nabla\cdot \bigl[(E + P^*) \mathbf{v} - \mathbf{B} (\mathbf{B \cdot v}) \bigr]&=&0,\label{eq:eoe}
\end{eqnarray}
where $P^* = p + \frac{B^2}{2}$ is the total pressure with $p$ being the gas pressure, $E = e + \frac{1}{2}\rho v^{2} + \frac{B^{2}}{2}$ is the total energy density with $e$ being the gas thermal energy, and $\mathbb I$ is the unit tensor. We use the ideal equation of state $p=(\gamma-1)e$ with the constant adiabatic index $\gamma$, or the barotropic (including isothermal) equation of state instead of the energy equation (\ref{eq:eoe}) depending on the problem.

All the simulations are performed using the Athena++ code \citep{athenapp}. Here we restrict our discussion to Cartesian coordinates for simplicity.

\subsection{Constrained Transport}
We briefly introduce the concept of the CT scheme here. While there are a few variants of the CT schemes, we use the method of \citet{athenact} implemented in Athena++ in this work. In this scheme, the magnetic fields $\mathbf{B}$ are defined at cell surfaces as area-averaged quantities, and the other physical variables such as $\rho, \mathbf{v}, E$ are defined at cell centers as volume-averaged quantities. The EMFs are calculated on cell edges using an appropriate combination of the EMFs at cell centers and fluxes estimated at cell surfaces from the Riemann solver. The magnetic fields at cell surfaces are updated by circular integration of the EMFs. Because the EMFs on each edge cancel out in this procedure, $\nabla\cdot\mathbf{B}$ of each cell is conserved. To calculate the thermal energy and gas pressure from the total energy, the magnetic fields at cell centers are necessary. They are calculated by a simple average of the surface fields.

\subsection{Mixed Divergence Cleaning}
\subsubsection{Formulation} \label{sec:formulation}
Here we review the mixed divergence cleaning method of \citet{dedner}. \citet{dedner} propose a few alternative formulae (Equations 24 and 38 in their paper). We focus on the basic GLM formula in the main part of this work because it is most commonly used in the public codes. We compare the GLM and Extended GLM (EGLM) formulae in Appendix~\ref{sec:eglm}.

Because it is important for understanding the behavior of the scheme, let us start from the foundation of the scheme. In this scheme, we introduce a new variable $\psi$ associated with $\nabla\cdot\mathbf{B}$, and modify the induction equation (\ref{eq:ind}) and solenoidal constraint (\ref{eq:sol}) like:
\begin{eqnarray}
\frac{\partial \mathbf{B}}{\partial t} - \mathbf{\nabla} \times \left(\mathbf{v} \times \mathbf{B}\right)+ \nabla \psi&=&0,\label{eq:ind2}\\
\mathcal{D}(\psi)+\nabla\cdot\mathbf{B}&=&0,\label{eq:sol2}
\end{eqnarray}
where $\mathcal{D}$ is a linear differential operator. From these equations, one can show that $\psi$ and $\nabla\cdot\mathbf{B}$ satisfy the same equation regardless of the choice of $\mathcal{D}$:
\begin{eqnarray*}
\frac{\partial \nabla\cdot\mathbf{B}}{\partial t}+\nabla^2\psi&=&0,\\
\frac{\partial \mathcal{D}(\nabla\cdot\mathbf{B})}{\partial t}+\nabla^2\mathcal{D}(\psi)&=&0,\\
\frac{\partial \mathcal{D}(\psi)}{\partial t}+\frac{\partial (\nabla\cdot\mathbf{B})}{\partial t}&=&0,\\
\nabla^2\mathcal{D}(\psi)+\nabla^2(\nabla\cdot\mathbf{B})&=&0,\\
\frac{\partial \mathcal{D}(\nabla\cdot\mathbf{B})}{\partial t}-\nabla^2(\nabla\cdot\mathbf{B})&=&0,\\
\frac{\partial \mathcal{D}(\psi)}{\partial t}-\nabla^2\psi&=&0.
\end{eqnarray*}
Therefore, $\psi$ can be used to cancel out $\nabla\cdot\mathbf{B}$. However, it is important to clarify the underlying assumptions here. First, the variables must be sufficiently smooth (differentiable). Second, the linear operator $\mathcal{D}$ must commute with the temporal and spatial derivative operators. As we will see later, the second assumption is not satisfied in the conventional implementation.

In the mixed divergence cleaning scheme, we adopt
\begin{eqnarray}
\mathcal{D}(\psi)=\frac{1}{c_h^2}\frac{\partial}{\partial t}\psi+\frac{1}{c_p^2}\psi.
\end{eqnarray}
Then, $\psi$ follows the telegraph equation, which is a mixture of the advection and diffusion equations:
\begin{eqnarray}
\frac{\partial^2 \psi}{\partial t^2}+\frac{c_h^2}{c_p^2}\frac{\partial \psi}{\partial t}-c_h^2\nabla^2\psi&=&0,
\end{eqnarray}
and the solenoidal constraint (\ref{eq:sol}) becomes:
\begin{eqnarray}
\frac{\partial \psi}{\partial t}+c_h^2\nabla\cdot\mathbf{B}&=&-\frac{c_h^2}{c_p^2}\psi.\label{eq:psi}
\end{eqnarray}
These equations mean that $\psi$ (hence the divergence error) is advected and diffuses out isotropically, where $c_h$ is the transport speed and $c_p^2$ is the diffusion coefficient. Alternatively, $c_h^2/c_p^2$ can be interpreted as the damping rate at which $\psi$ declines. Then, the $\nabla\psi$ term in eq.(\ref{eq:ind2}) is supposed to cancel out the divergence error in the magnetic fields.

\subsubsection{The Conventional Implementation}\label{sec:implementation}
In this scheme, all the variables including the magnetic fields are defined as cell-centered variables. In many modern MHD codes, the system is solved using approximate Riemann solvers. We solve the hyperbolic part and source term on the right-hand side in eq.(\ref{eq:psi}) separately using the operator splitting technique. The subsystem of the normal component of the magnetic field (e.g. $B_x$) and $\psi$ are decoupled from the other MHD equations, and can be solved using the numerical flux presented in \citet{dedner}. In the $x$-direction, for example, $B_x$ and $\psi$ at the interface between the $i$-th and $i+1$-th cells are calculated as
\begin{eqnarray}
B_{x,i+1/2,j,k}&=&\frac{1}{2}\left(B_{x,l}+B_{x,r}\right)-\frac{1}{2c_h}\left(\psi_{r}-\psi_{l}\right),\label{eq:bxm}\\
\psi_{i+1/2,j,k}&=&\frac{1}{2}\left(\psi_l+\psi_r\right)-\frac{c_h}{2}\left(B_{x,r}-B_{x,l}\right),\label{eq:psim}
\end{eqnarray}
where the values with subscript $l$ and $r$ are the values on the left and right sides of the interface calculated by a reconstruction method. Then the cell-centered variables are updated using the fluxes calculated as follows.
\begin{eqnarray}
F_{B_x,i+1/2,j,k}&=&\psi_{i+1/2,j,k},\label{eq:bxflux}\\
F_{\psi,i+1/2,j,k}&=&c_h^2 B_{x,i+1/2,j,k}.\label{eq:psiflux}
\end{eqnarray}
The source term,
\begin{eqnarray}
\frac{\partial \psi}{\partial t}&=&-\frac{c_h^2}{c_p^2}\psi,\label{eq:src}
\end{eqnarray}
is solved separately using the operator-splitting technique, which allows us to solve it exactly if $\frac{c_h^2}{c_p^2}$ remains constant:
\begin{eqnarray}
\psi^{n+1}=\exp\left(-\frac{c_h^2}{c_p^2}\Delta t_{n}\right)\psi^*,\label{eq:damp}
\end{eqnarray}
where $\Delta t_n$ is the timestep at the $n$-th step and $\psi^*$ is the intermediate value after applying the hyperbolic part.

\subsubsection{Transport Speed and Diffusion Coefficient} \label{sec:ch}
The scheme contains two arbitrary parameters: the transport speed $c_h$ and the diffusion coefficient $c_p^2$. These parameters must be chosen carefully so that the scheme remains stable, accurate, and computationally efficient. In the conventional implementation, we use the largest possible $c_h$ which does not introduce any additional timestep constraint:
\begin{eqnarray}
c_h=\frac{\eta_{\rm CFL} h_{\rm min}}{\Delta t_n} \label{eq:ch}
\end{eqnarray}
where $\eta_{\rm CFL}$ is the Courant-Friedrichs-Levy (CFL) number in use for the MHD part to control the timestep, and $h_{\rm min}$ is the minimum cell spacing among all the cells and directions. If the timestep $\Delta t_n$ is determined by the fastest signal speed in the system $c_{\rm max}$, or $\Delta t_n = \frac{\eta_{\rm CFL} h_{\rm min}}{c_{\rm max}}$, it simply means $c_h = c_{\rm max}$.

The choice of $c_p$ is more arbitrary. \citet{dedner} suggest using the ratio between the hyperbolic and parabolic coefficients as the parameter\footnote{As an alternative, they also suggest to fix the damping factor in eq.(\ref{eq:damp}) : $c_d\equiv\exp\left(-\frac{c_h^2}{c_p^2}\Delta t\right)$.}: $c_r\equiv c_p^2/c_h$. They claim that the scheme's behavior is not very sensitive to it (Figure 2 of \citet{dedner}), and they recommend $c_r=0.18$.

There are a few issues in their formulation. First, $c_r$ is not dimensionless, and therefore depends on units \citep{price05}. In order to clarify its meaning, we should redefine it as the ratio between the advection and diffusion timescales at the scale of $L$:
\begin{eqnarray}
C_r\equiv\frac{t_{\rm adv}}{t_{\rm diff}}=\frac{L/c_h}{L^2/c_p^2}=\frac{c_p^2}{c_h L},\label{eq:cr}
\end{eqnarray}
where $C_r$ is dimensionless. Using this, the source term (\ref{eq:src}) can be rewritten as:
\begin{eqnarray*}
\frac{\partial \psi}{\partial t}&=&-\frac{c_h}{C_r L}\psi=-\frac{1}{C_r t_{\rm adv}}\psi.
\end{eqnarray*}
The original formula corresponds to setting $L=1$, but the choice of the scale is arbitrary\footnote{Changing $L$ is essentially equivalent to changing $C_r$. However, because almost all the implementations use $C_r=0.18$ as suggested in the original paper, we fix it and change $L$ instead in this work.}. Some simulation codes such as PLUTO \citep{pluto,mignone10} for example adopt $L=h_{\rm min}$. With adaptive mesh refinement (AMR), the diffusion (or damping) timescale may change in such an implementation. As it is not clear how the choice of this parameter affects solutions, we compare them in Section~\ref{sec:results}. 
It should be noted that $\tau_{\rm d} \equiv C_r t_{\rm adv}$ is the damping timescale, and $C_r$ is the ratio between the damping timescale and crossing timescale $t_{\rm adv}$\footnote{It should be also noted that the ratio between the damping time and the timestep is independent of $\Delta t_n$; $\frac{\tau_{\rm d}}{\Delta t_n}=\frac{C_r}{\eta_{\rm CFL}}\frac{L}{h_{\rm min}}$. Therefore, while the damping time is well resolved if $L\gg h_{\rm min}$, it is poorly resolved if $L\sim h_{\rm min}$. In the latter case, an implicit or exact time integrator like eq.(\ref{eq:damp}) is necessary. We also tested the first-order backward Euler integrator, but the results remain qualitatively similar although the backward Euler integrator gives weaker (slower) damping.}. In other words, during the damping timescale, the divergence cleaning variable $\psi$ propagates over the distance of $C_r L$, indicating that the divergence error is processed more locally with smaller $L$.

Second, eq.(\ref{eq:ch}) means that the transport speed is uniform in space but not constant in time\footnote{While most codes adopt the spatially-uniform transport speed, \citet{tricco16} and \citet{gizmo} use the fastest local characteristic speed as $c_h$ which can be spatially non-uniform. Also, grid based codes with mesh refinement and individual timestepping may use spatially non-uniform $c_h$ if it is calculated according to the local timestep. In such implementations, the differential operator $\mathcal{D}$ does not commute with the spatial derivative, which may produce artificial structures where $c_h$ changes sharply.} because it depends on the timestep $\Delta t_n$. Because the timestep is usually set from the MHD part (and other additional physical processes if necessary), this means that it violates the second assumption discussed in Section~\ref{sec:formulation}; the differential operator $\mathcal{D}$ does not commute with the time derivative\footnote{It should be noted that changing $c_p$ during a simulation is also inconsistent with the derivation. For example, use of the constant damping factor $c_d$ and/or $L=h_{\rm min}$ with adaptive mesh refinement can violate the assumption. However, it seems less problematic than changing $c_h$, possibly because $c_p$ appears only in the damping term and does not directly affect the fluxes nor energies.}. As demonstrated later, this causes an unphysical behavior when the timestep changes drastically.

In order to mitigate the issue, a fixed value of $c_h$ should be used instead of changing it depending on the timestep \citep{arepo}. In order for the scheme to be stable, $c_h$ must be carefully selected so that it is larger than a certain characteristic speed in the system. Then, the timestep must be adjusted to satisfy the CFL condition due to this transport speed. While this implementation can suppress the unphysical behavior caused by a sudden change in the timestep, there are, however, several practical difficulties with this implementation. It is not trivial to estimate the maximum speed before actually running a simulation. Moreover, the additional computational cost due to the smaller timestep can be enormous, particularly when the fluid and Alfv\'{e}n speed changes drastically during a simulation. Star formation in a collapsing cloud and galactic dynamo are notable examples of such a system where the magnetic fields are amplified by gravitational collapse and differential rotation.

\section{Test Calculations} \label{sec:results}
We compare the CT scheme and divergence cleaning scheme through a few test problems in this section. We label our models as in Table~\ref{tab:methods}. We adopt the same simulation configurations for everything else; we adopt the second-order van-Leer time integrator, second-order piecewise linear method (PLM) using primitive variables for spatial reconstruction, and the Harten-Lax-van Leer Discontinuities (HLLD) approximate Riemann solver \citep{hlld}.

First, we compare the schemes using a series of Kelvin-Helmholtz instability (KHI) simulations in two dimensions. These test calculations are intended to clarify how the schemes behave in simple setups and provide guidance for the choice of the parameters in the divergence cleaning method. Then we present comparisons based on more practical applications of collapsing clouds and atmospheric convection.

\begin{deluxetable}{llll}
\tablecaption{ \label{tab:methods}}
\tablehead{\colhead{Model} & \colhead{Solver} & \colhead{$L$} & \colhead{$c_h$}}
\startdata
CT & Constrained Transport & --- & --- \\
D1v & Mixed $\nabla\cdot\mathbf{B}$ Cleaning & $L=1$ & variable (\ref{eq:ch}) \\
Dhv & Mixed $\nabla\cdot\mathbf{B}$ Cleaning & $L=h$ & variable (\ref{eq:ch}) \\
D1c & Mixed $\nabla\cdot\mathbf{B}$ Cleaning & $L=1$ & constant \\
Dhc & Mixed $\nabla\cdot\mathbf{B}$ Cleaning & $L=h$ & constant \\
\enddata
\tablecomments{The third column indicates the scale $L$ used in eq.(\ref{eq:cr}), with $h$ being the finest resolution. The fourth column is the transport speed $c_h$, and the adopted constant values are specified in each test problem.}
\end{deluxetable}

\subsection{KHI with a Uniform Magnetic Field}\label{sec:khuni}

\begin{figure*}[t!]
\includegraphics[bb=0 0 759.105959 892.624596,width=\textwidth]{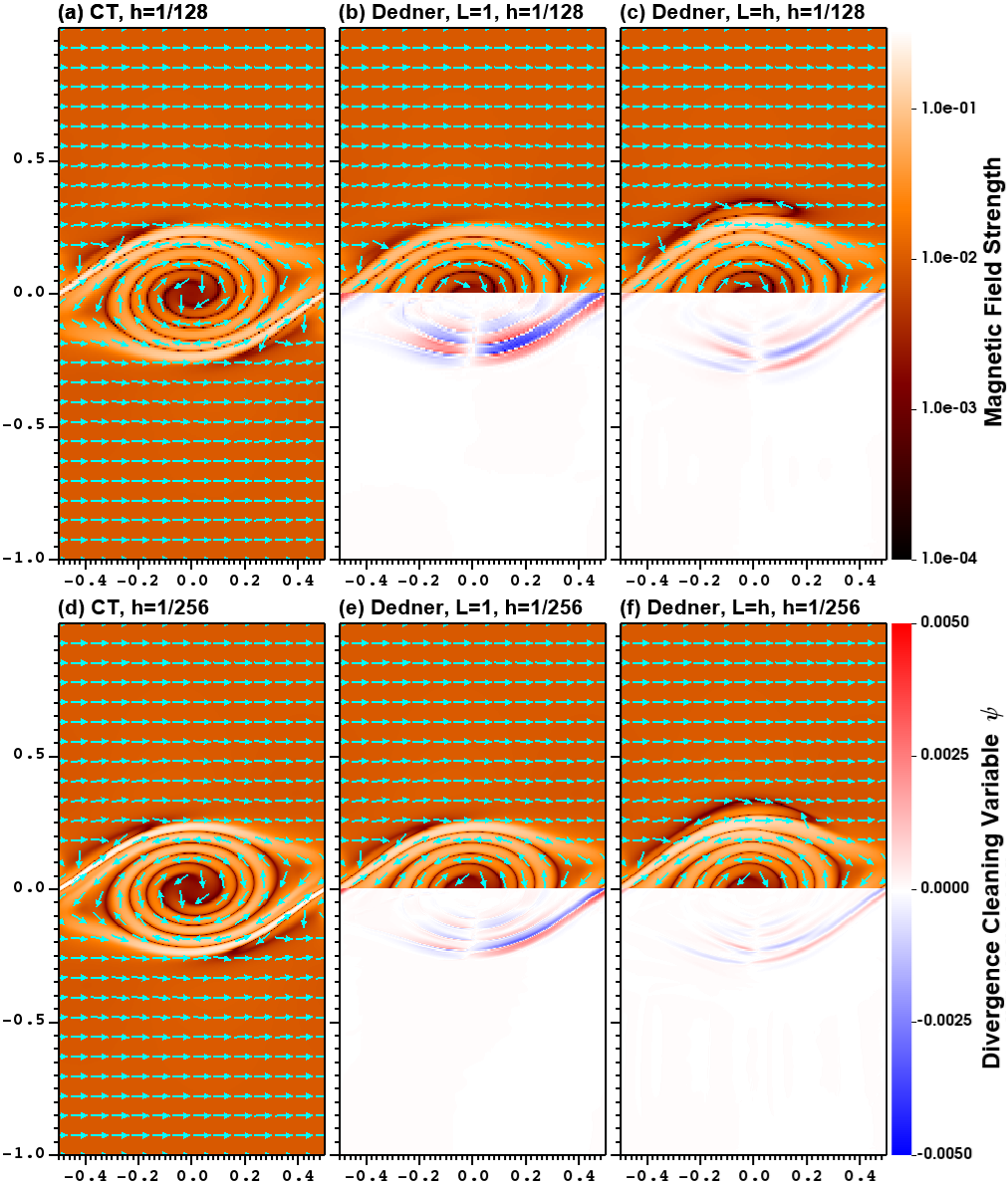}
\caption{Distributions of the magnetic field strength with magnetic field directions (orange color maps with cyan arrows) and the divergence cleaning variable (shown in the lower halves of panels (b), (c), (e) and (f)) in the KHI test with uniform magnetic fields at $t=3$. Panels (a) and (d) correspond to CT, (b) and (e) to D1v, and (c) and (f) to Dhv. The top and bottom rows show the low- ($h=1/128$) and high-resolution ($h=1/256$) models, respectively. \label{fig:kh_uniform}}
\end{figure*}

\begin{figure*}[t!]
\includegraphics[bb=0 0 1117.656006 645.090044,width=\textwidth]{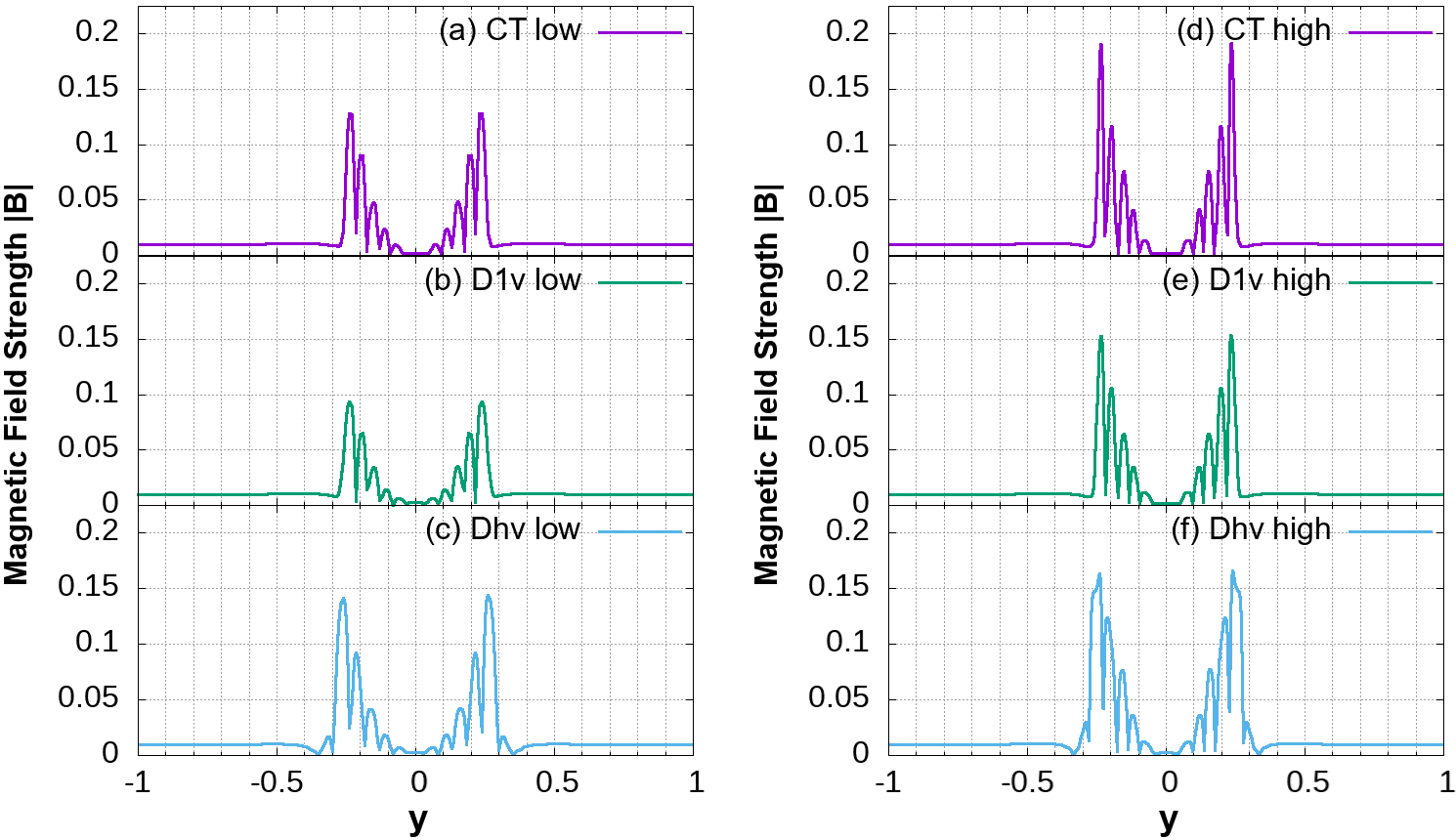}
\caption{Profiles of the magnetic field strength at $x=0$ in the KHI test with uniform magnetic fields at $t=3$. The top, middle, and bottom rows show CT, D1v and Dhv models, and the left and right columns show the low- and high-resolution models, respectively. The labels (a) -- (f) correspond to those in Figure~\ref{fig:kh_uniform}. \label{fig:kh_profu}}
\end{figure*}

\begin{figure*}[t!]
\includegraphics[bb=0 0 970.635485 892.624596,width=\textwidth]{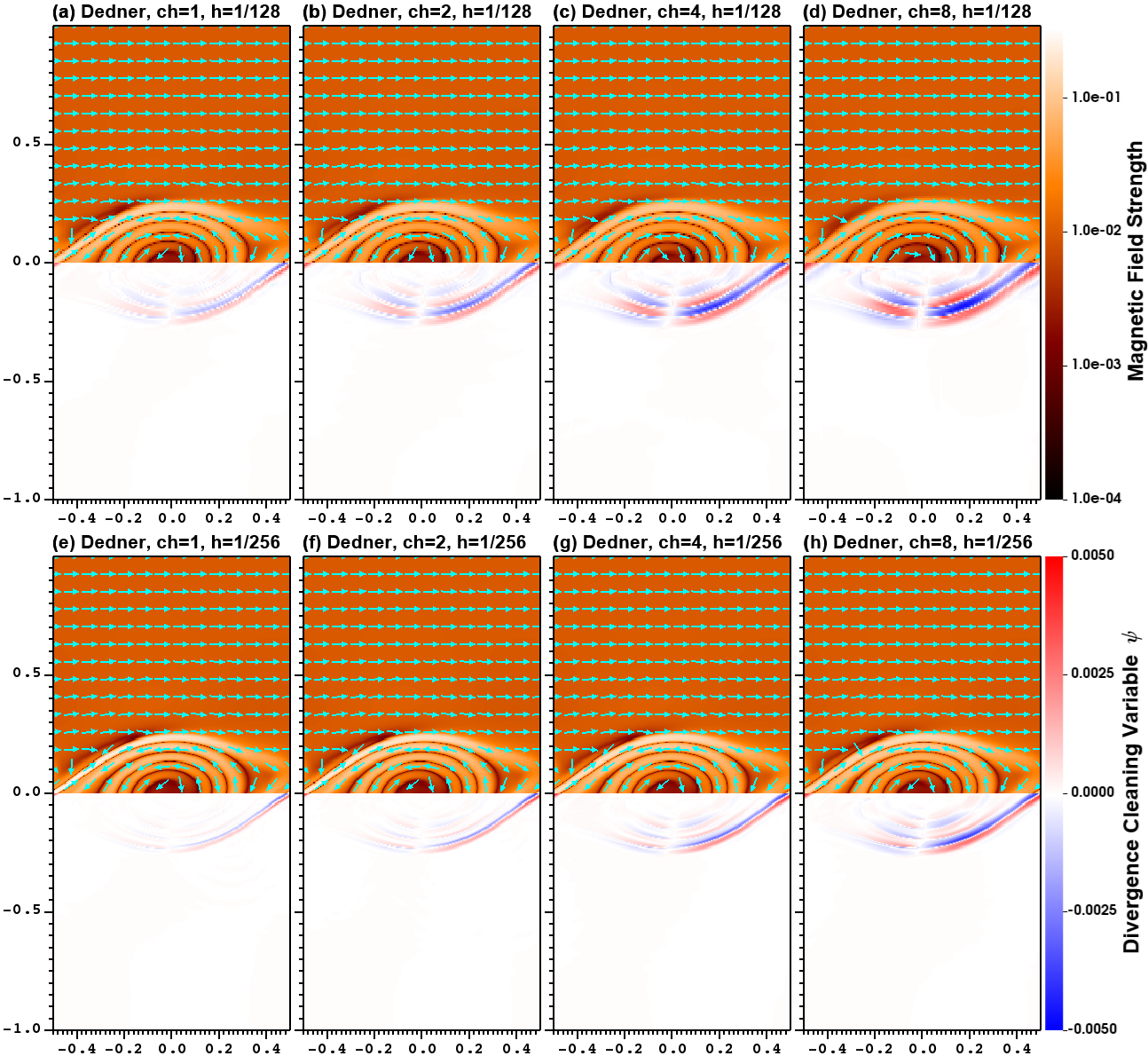}
\caption{Comparison between the D1c models with different $c_h$. From left to right, $c_h=1.0, 2.0, 4.0, 8.0$. The top and bottom rows show the low- ($h=1/128$) and high-resolution ($h=1/256$) models, respectively. \label{fig:kh_ch}}
\end{figure*}

\begin{figure*}[t!]
\includegraphics[bb=0 0 759.105959 892.624596,width=\textwidth]{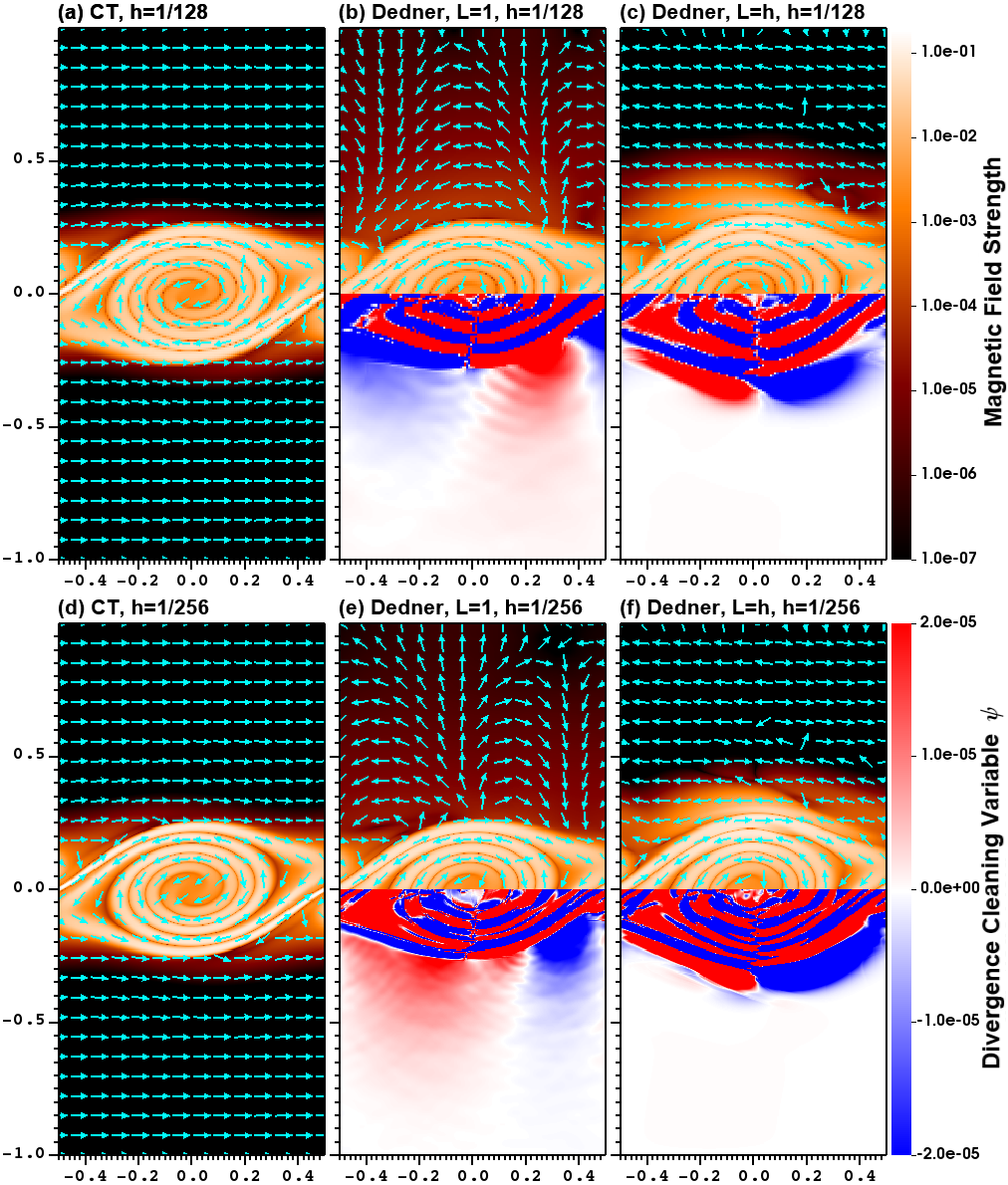}
\caption{Same as Figure~\ref{fig:kh_uniform} but for the KHI test with localized magnetic fields. Note that the color bar ranges are different. The color range for the divergence cleaning variable is set intentionally narrow to show the structure in the weakly-magnetized region. \label{fig:kh_localized}}
\end{figure*}

\begin{figure*}[t!]
\includegraphics[bb=0 0 1120.656425 645.090044,width=\textwidth]{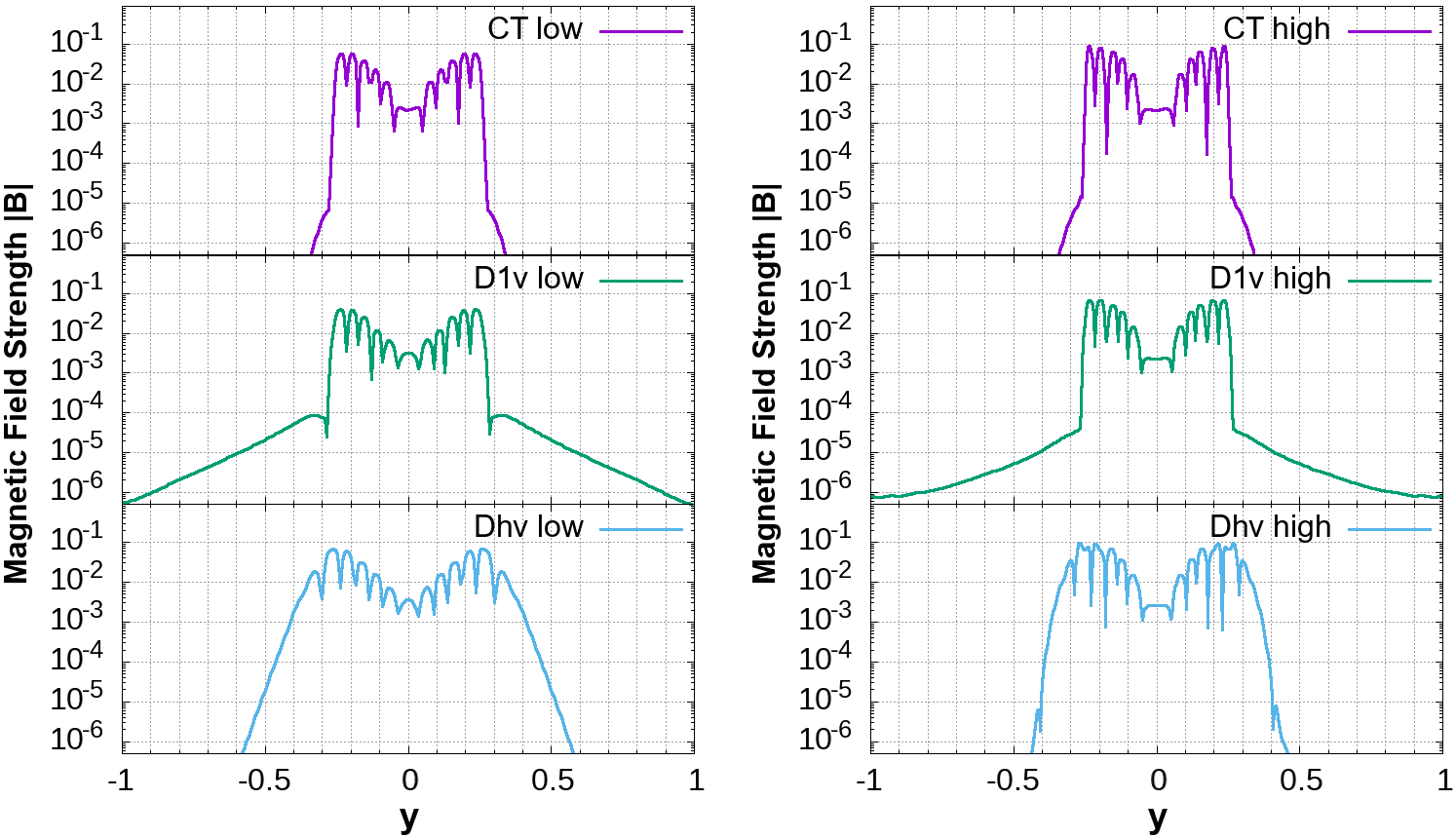}
\caption{Same as Figure~\ref{fig:kh_profu} but for the KHI test with localized magnetic fields. Note that the vertical axes are logarithmic. \label{fig:kh_profloc}}
\end{figure*}

We first perform a KHI test with a uniform magnetic field in two dimensions. The computational domain spans [-0.5:0.5] $\times$ [-1.0:1.0]. To investigate the convergence behaviors, we run the simulations with $\Delta x = \Delta y = h = 1/128$ and $1/256$. We initialize the problem with a uniform density $\rho=1$, pressure $p=10$ and magnetic field $\mathbf{B}=(0.01,0,0)$, and the adiabatic index $\gamma=5/3$. The corresponding plasma beta is $\beta=2\times 10^5$, meaning that the initial magnetic fields are weak and dynamically unimportant. For the models with divergence cleaning, we also set $\psi=0$ uniformly. The velocity is initialized as
\begin{eqnarray*}
v_x&=&v_0\tanh\left(\frac{y}{a}\right),\\
v_y&=&A\sin\left(2\pi x\right)\exp\left(-\frac{y^2}{\sigma^2}\right),\\
v_z&=&0,
\end{eqnarray*}
where $v_0=1.0$ is the shearing velocity, $a=0.05$ the thickness of the layer, $A=0.01$ the amplitude of the initial perturbation, and $\sigma=0.2$ the thickness of the perturbation, respectively. We use the CFL number $\eta_{\rm CFL}=0.4$ in this test. This subsonic shearing layer is well resolved, and there is a single mode in the $x$-direction. The boundary conditions are periodic in the $x$-direction and zero gradient (outflow) in the $y$-direction. The magnetic fields are initially weak and evolve passively, but are amplified by the vortex motion as the instability develops. We compare the results at t = 3.0 as significant resolution dependence arises beyond this point because of numerical reconnections and nonlinear behaviors of the system.

\subsubsection{CT vs Conventional Divergence Cleaning}\label{sec:ct_vs_dedner}

We present the results in Figures~\ref{fig:kh_uniform} and \ref{fig:kh_profu}. We first compare CT and the conventional implementation of the divergence cleaning with the variable transport speed $c_h$ in eq.(\ref{eq:ch}), based on the low resolution models in panels (a)--(c). Overall, the solutions in CT (a) and D1v (b) are qualitatively consistent, although the amplitude of the magnetic fields is higher in CT, indicating that CT is less diffusive. In contrast, the solution of Dhv (c) is qualitatively different. The vortex is significantly more expanded vertically, and magnetic field bumps emerge above and below the vortex around $y\sim\pm 0.3$. These structures do not appear in the other models and seem to be artifacts due to the solver. Although the magnetic field amplitude is higher in Dhv, the solution of D1v is more consistent with CT.

The same trends hold in the high resolution models in panels (d)--(f). The solutions obtained with CT (a) and (d) are almost converged with the stronger magnetic fields. In the divergence cleaning schemes, the distributions of $\psi$ in (e) and (f) are more localized with reduced amplitude in the higher resolution models. Comparing D1v and Dhv, D1v behaves more consistently with the higher maximum field strength (e). On the other hand, Dhv exhibits more significant resolution dependency (f). While the field amplitude slightly increases as expected, the vortex is less expanded vertically and the magnetic bumps around $y\sim\pm0.3$ persist in the high resolution case. The divergence cleaning variable $\psi$ becomes smaller and more localized, indicating that the higher resolution models produce less divergence errors, as expected.

These results indicate that CT tends to be less dissipative than the divergence cleaning schemes, and the choice of the parameters for the divergence cleaning methods has a significant impact on solutions.

\subsubsection{Divergence Cleaning with Constant Transport Speed}
Next, we compare the behaviors of the divergence cleaning schemes with constant transport speeds in Figure~\ref{fig:kh_ch}. As we already know that the scheme with $L=h$ performs somewhat inconsistently, here we focus on the models with $L=1$ (D1c). First, we find that the system is numerically unstable for a small transport speed ($c_h \lesssim 1.0$)\footnote{Numerical instabilities develop rapidly with $c_h=0.8$ or lower in the low resolution models with $h=1/128$. The system seems to be on the verge of stability at $c_h=0.9$, as it remains qualitatively stable up to $t=50$ but ripples with small amplitudes are visible occasionally. A similar trend is observed in the high resolution models with $h=1/256$, indicating that this stability threshold is not sensitive to the resolution.}. Beyond that, the solutions become more diffusive with higher transport speeds; the magnetic field layer gets thicker and the field amplitude decreases. This is expected because the divergence error propagates further with the larger transport speed. The difference due to the different transport speeds becomes less pronounced in the higher resolution models. Note that the models discussed in Section~\ref{sec:ct_vs_dedner} correspond to $c_h\sim 4$, although it gradually increases as the system evolves.

\begin{figure*}[t!]
\begin{center}
\includegraphics[bb=0 0 1393.694537 892.624596,angle=90,width=0.8\textwidth]{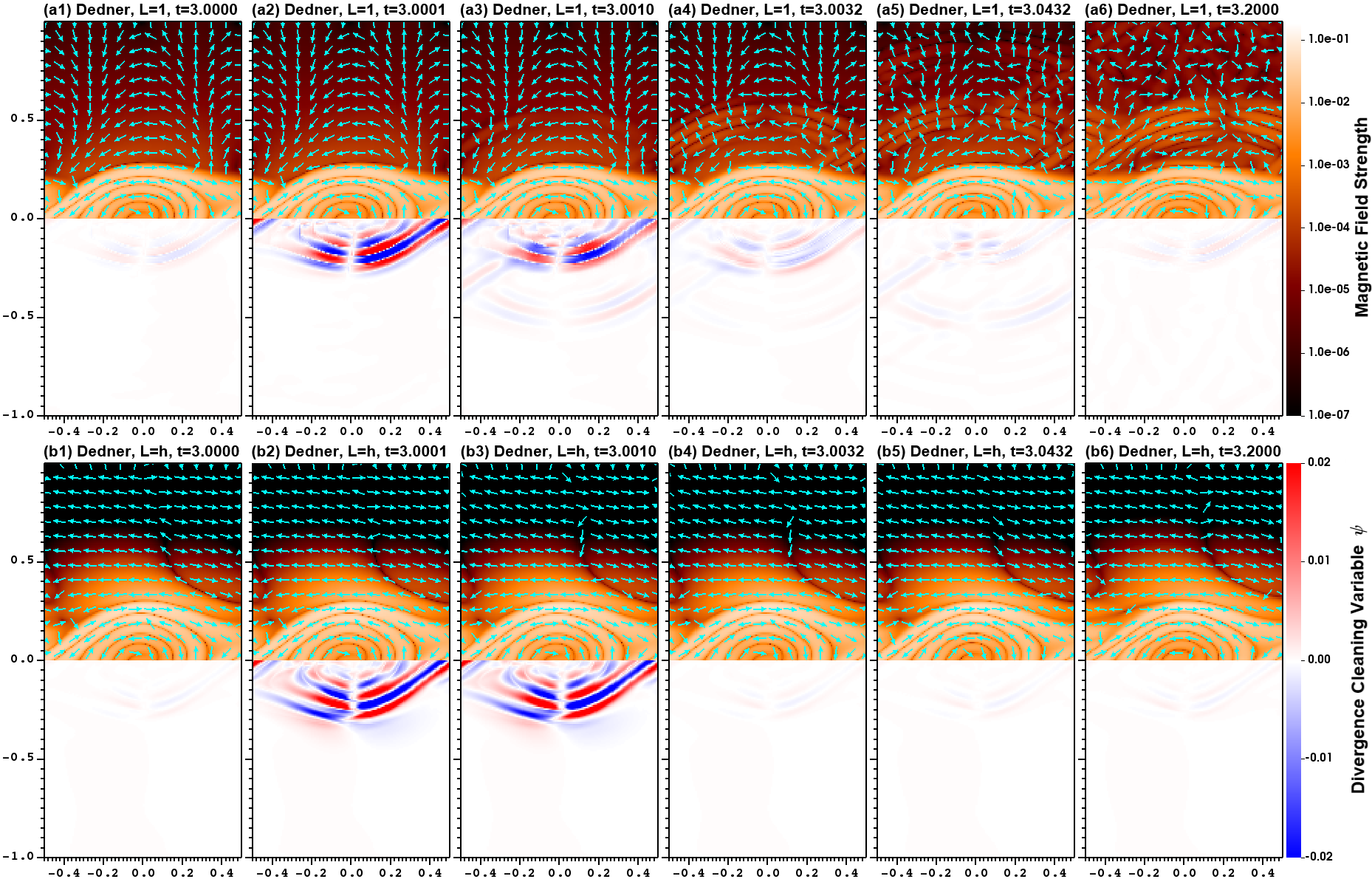}
\end{center}
\caption{Time evolution of the magnetic field strength with magnetic field directions (orange color maps with cyan arrows) and the divergence cleaning variable (bottom half) in the D1v (top) and Dhv (bottom) models before and after the sudden change in the timestep.\label{fig:kh_dt1}}
\end{figure*}

\begin{figure}[t!]
\begin{center}
\includegraphics[bb=0 0 767.904012 575.928009,width=\hsize]{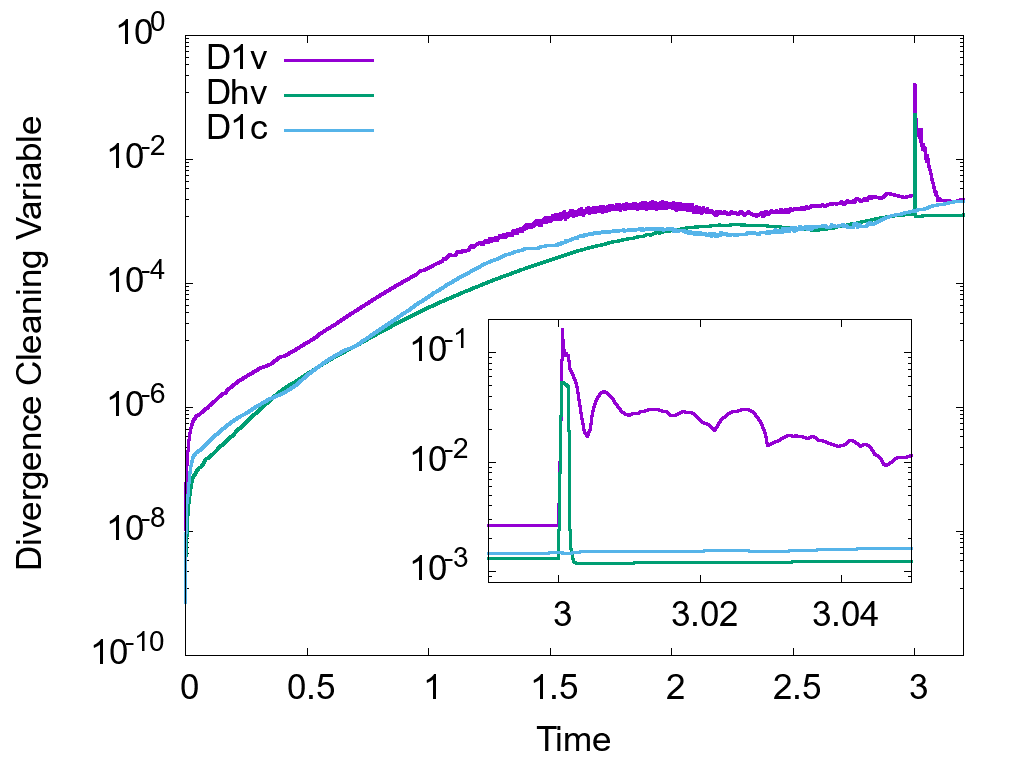}
\end{center}
\caption{Time evolution of the maximum divergence cleaning variable.\label{fig:psimax}}
\end{figure}

\begin{figure*}[t!]
\begin{center}
\includegraphics[bb=0 0 1393.694537 892.624596,angle=90,width=0.8\textwidth]{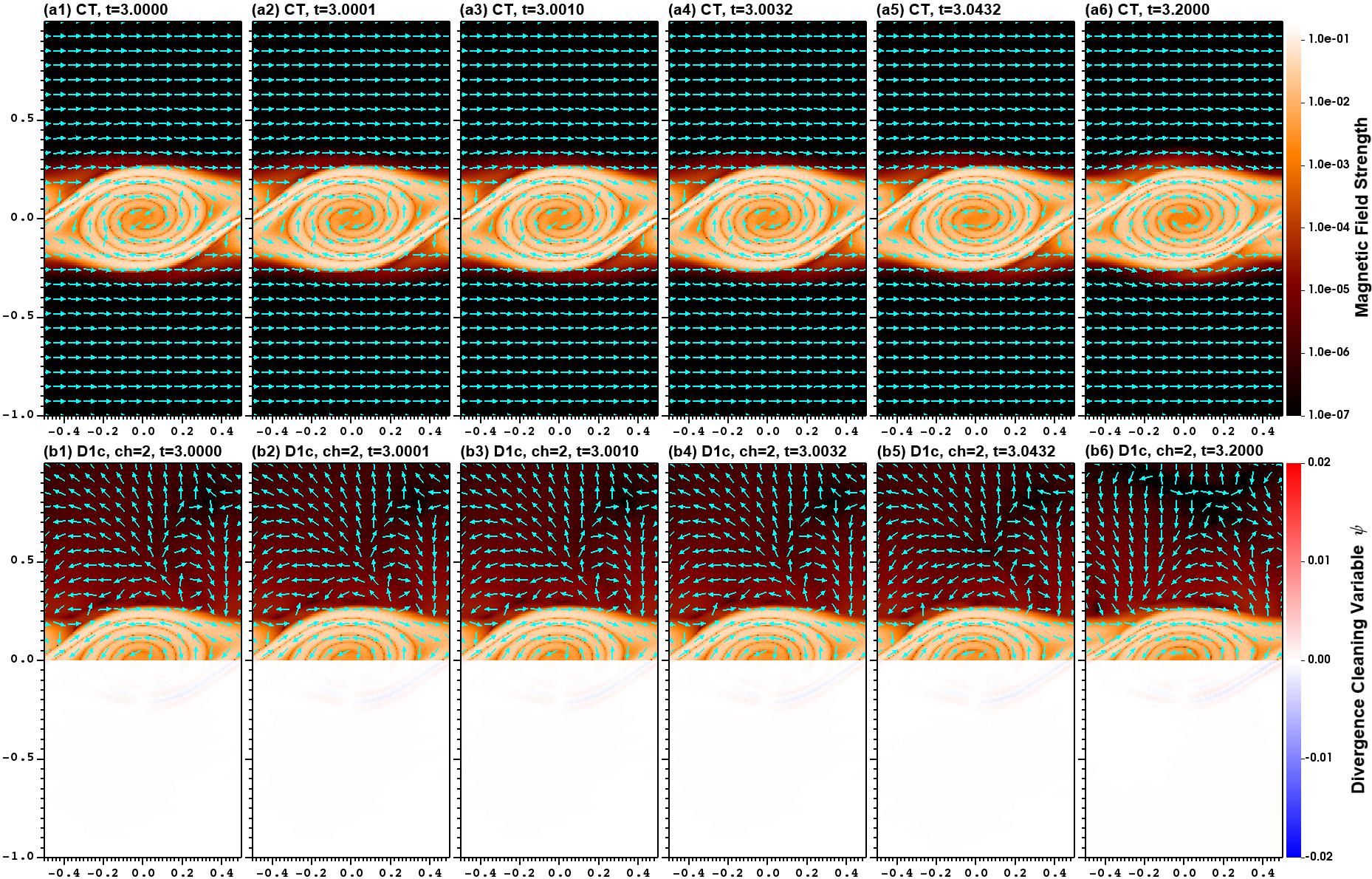}
\end{center}
\caption{Same as Figure~\ref{fig:kh_dt1} but for CT and D1c. \label{fig:kh_dt2}}
\end{figure*}

\subsection{KHI with a Localized Magnetic Field}\label{sec:khloc}
In the mixed divergence cleaning scheme, the divergence error is transported isotropically while being damped. This means that the divergence error is not locally processed at the generation point. This raises a concern that, when magnetic fields are strongly localized, the divergence error produced in the strongly magnetized region can propagate to the weakly magnetized region and strongly perturb the magnetic field there.

In order to investigate the behavior of the schemes in such a situation, we perform another set of KHI simulations with localized magnetic fields. We initialize the magnetic field with a Gaussian distribution around the shearing layer:
\begin{equation}
B_x=B_0\exp\left(-\frac{y^2}{\sigma_2^2}\right)+B_{\rm min},\\
\end{equation}
where $B_0=0.01$ is the maximum magnetic field strength and $\sigma_2=0.1$ the thickness of the magnetized region, respectively. The minimum magnetic field of $B_{\rm min}=10^{-30}$ is added in order to give a well-defined direction to the magnetic field even in the very weakly magnetized region, but this is just for plotting and is dynamically unimportant. The other variables remain the same as in the previous test.

The results are shown in Figures~\ref{fig:kh_localized} and \ref{fig:kh_profloc}. The evolution of the vortex remains qualitatively the same as in the previous test, because the magnetic fields are initially weak and evolve passively. In the weakly magnetized region outside the vortex, significant differences arise between the schemes. CT can maintain the magnetic fields localized, and the weakly magnetized region remains almost unaffected. However, the divergence cleaning schemes, both with $L=1$ and $L=h$, cannot maintain the magnetic fields confined. Just outside the vortex ($y\sim 0.3$), the amplitudes of the leaked magnetic fields reach $|B|\sim 10^{-4}$ in D1v and $|B|\sim 10^{-2}$ in Dhv, although it remains more localized in the latter. This is expected because the divergence error is processed more locally with smaller $L$, as discussed in Section~\ref{sec:ch}. The geometries of the leaked fields are different from the initial field, and not physical. In D1v, the leaked fields exhibit arch-like shapes (panels (b) and (e)). In Dhv, the field vectors are pointing more or less to the $x$ direction in the top-right part, but they are pointing opposite in the top-left part. These leaked fields do not satisfy the solenoidal constraint in some regions, because they are produced by the gradient of $\psi$. The trend remains similar even in the high resolution models, as well as in models with constant transport velocities. It should be noted that the field geometries in the weakly magnetized regions are different between the models with different resolutions, indicating that these artificially produced fields do not behave consistently even with higher resolutions.

\subsection{KHI with a Sudden Timestep Change}\label{sec:khdt}
We also point out in Section~\ref{sec:ch} that the divergence scheme is inconsistent with its derivation if the timestep is not constant. However, in the conventional implementation suggested in the original paper \citep{dedner}, the transport speed in eq.(\ref{eq:ch}) changes according to the timestep. In practical applications, the timestep can change as the system evolves. Usually, the timestep changes gradually as long as the system evolves continuously. However, certain situations like numerical glitches or additional physics such as fast radiation cooling can cause abrupt and drastic timestep changes in practical applications. Here we demonstrate that the conventional divergence cleaning method does not behave consistently in such a situation.

We use the same test problem as in the previous section. We use the fixed timestep of $\Delta t=4.0\times 10^{-4}$ instead of using a constant CFL number until $t=3.0$. We then decrease it to $\Delta t=1.0\times 10^{-5}$ for 100 steps, and calculate until $t=3.2$ with $\Delta t=4.0\times 10^{-4}$ again\footnote{Athena++ gradually increases $\Delta t$ even when a large timestep is allowed, while it immediately decreases it when a small timestep is required. Although we do not have to use this feature in this test, we keep this as it reflects practical use cases.}.

The results with the conventional divergence cleaning are presented in Figure~\ref{fig:kh_dt1}. We also show the evolution of the maximum divergence cleaning variable $\psi$ in Figure~\ref{fig:psimax}. Before the timestep changes (panels (a1) and (b1) in Figure~\ref{fig:kh_dt1}), the maximum divergence cleaning variable remains more or less constant, indicating that generation of the divergence error is balanced with the cleaning. When the short timestep is specified at $t=3.0$, $\psi$ gets amplified drastically, both in D1v and Dhv. It quickly reaches a maximum value at $\sim 40$ times larger than the value before the timestep change\footnote{From additional calculations, we find that the amplitude of the jump in the divergence is almost proportional to the ratio of the timestep change.} in less than 10 timesteps indicating that the system quickly settles down to a new balanced state (panels (a2) and (b2)). In the next 100 steps with $\Delta t=1.0\times 10^{-5}$, ripple-like patterns are generated in the vortex and propagate to the outer region, and produce some visible (but not drastic) perturbation in the magnetic fields in D1v (a3). Because these ripples propagate at a speed of $c_h$ which is largely enhanced by the short timestep, they propagate rapidly and affect a large area. Once the timestep returns to the original value, while $\psi$ declines back to an amplitude similar to that before the timestep change, prominent ripples appear in the magnetic fields (a4). The weakly-magnetized region is completely dominated by this spurious field produced by this effect, which propagates again at the speed of $c_h$. Because these fields are produced numerically by the $\nabla\psi$ term in the modified induction equation (\ref{eq:sol2}), they carry a high level of the divergence error. The vortex continues to generate the ripples (a5), and the system does not recover its original state even after $\sim 500$ steps (a6). On the other hand, Dhv seems to be less susceptible to this effect. Even though $\psi$ is amplified, it quickly goes back to the original level once the long timestep is recovered (b3 -- b6). 

The timestep change does not affect CT and the divergence cleaning with a constant transport speed (D1c) as shown in Figure~\ref{fig:kh_dt2}. For these schemes, the timestep change only affects the truncation error in the time integrator, which is usually minor in an explicit scheme with an integrator of second order or higher. These results clearly indicate that we should use an appropriate constant transport speed $c_h$. It should be noted, however, that the strongly localized magnetic fields still leak into the weakly magnetized region even with the constant transport speed.

\subsection{Magnetic Field Amplification in a Collapsing Cloud}\label{sec:collapse}
In order to compare the behaviors of the schemes in a practical application, we perform ideal MHD simulations of gravitational collapse of rotating clouds with very weak initial magnetic fields. This is motivated by \citet{hirano22} and \citet{machida25} who studied rapid magnetic field amplification in the context of the Pop-III star formation.

\begin{figure*}[t!]
\begin{center}
\includegraphics[bb=0 0 1189.666058 1797.250866,width=0.82\hsize]{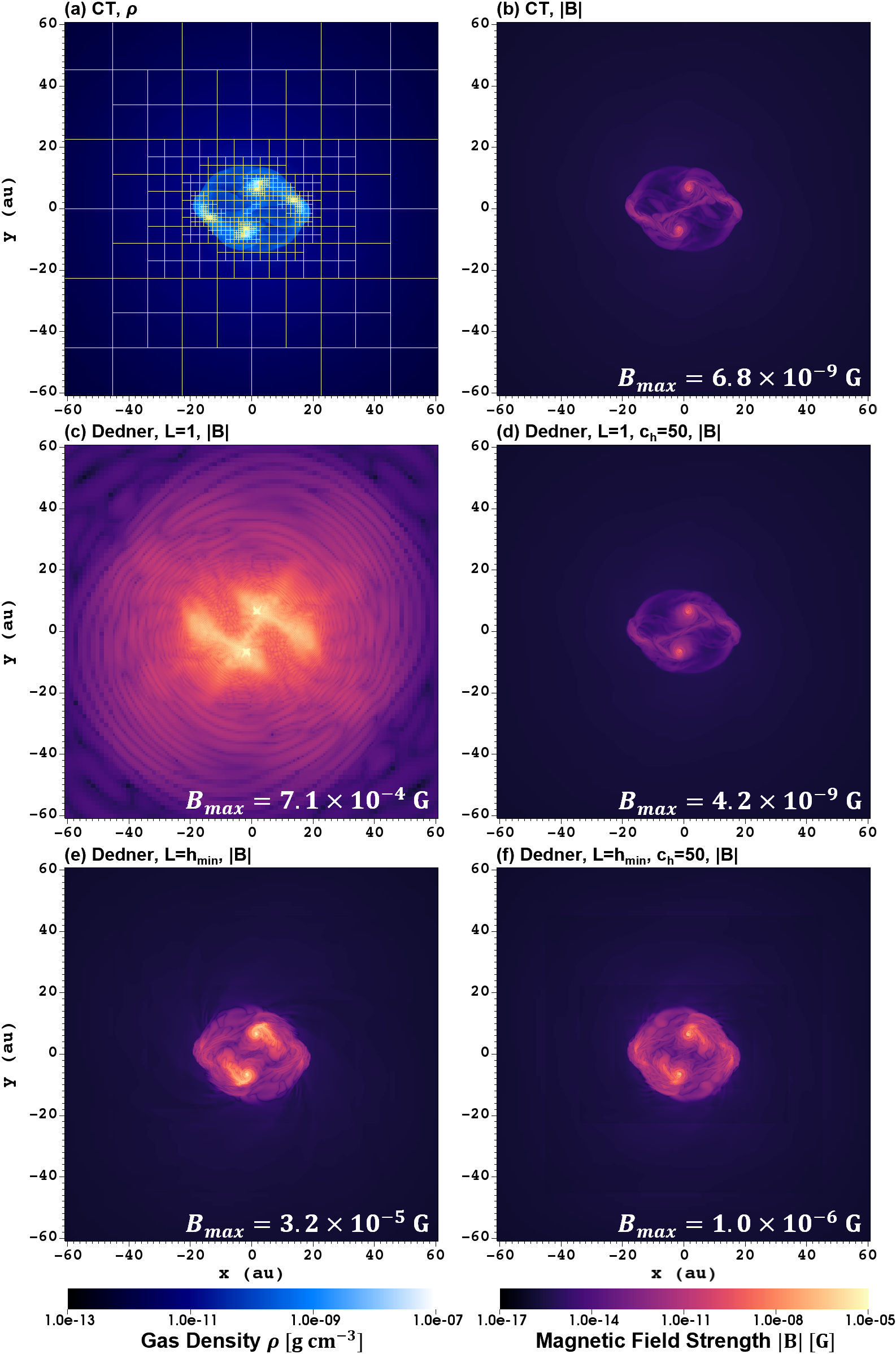}
\end{center}
\caption{Results of the simulations of the collapsing clouds in the early Universe. (a) Density cross section with MeshBlock boundaries of CT, (b)--(f) Magnetic field strength of CT, D1v, D1c with $c_h=50$, Dhv, Dhc with $c_h=50$, respectively. The maximum magnetic field strength is shown on each panel. The different sizes of the MeshBlocks in (a) correspond to different AMR refinement levels. \label{fig:collapse}}
\end{figure*}

\begin{figure*}[t!]
\begin{center}
\includegraphics[bb=0 0 1332.185951 1143.159566,width=\hsize]{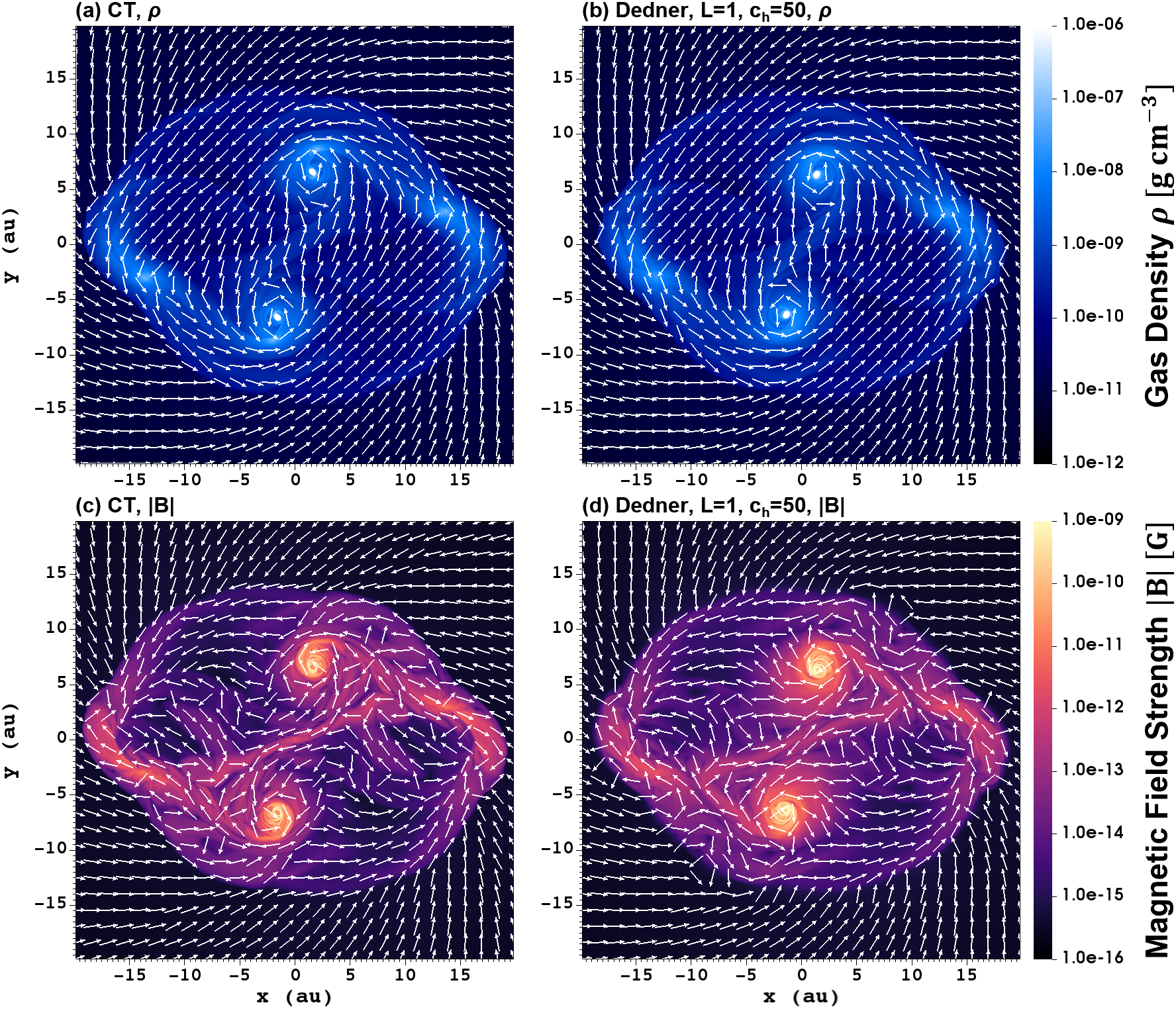}
\end{center}
\caption{Zoom-in views of the collapsing clouds in the early Universe. Panels (a) and (b) show density cross sections with gas velocity directions for CT and D1c with $c_h=50$, respectively, while panels (c) and (d) show magnetic field strength with magnetic field directions for the same models. \label{fig:collapsez}}
\end{figure*}

\begin{figure}[htb]
\begin{center}
\includegraphics[bb=0 0 767.904012 575.928009,width=\hsize]{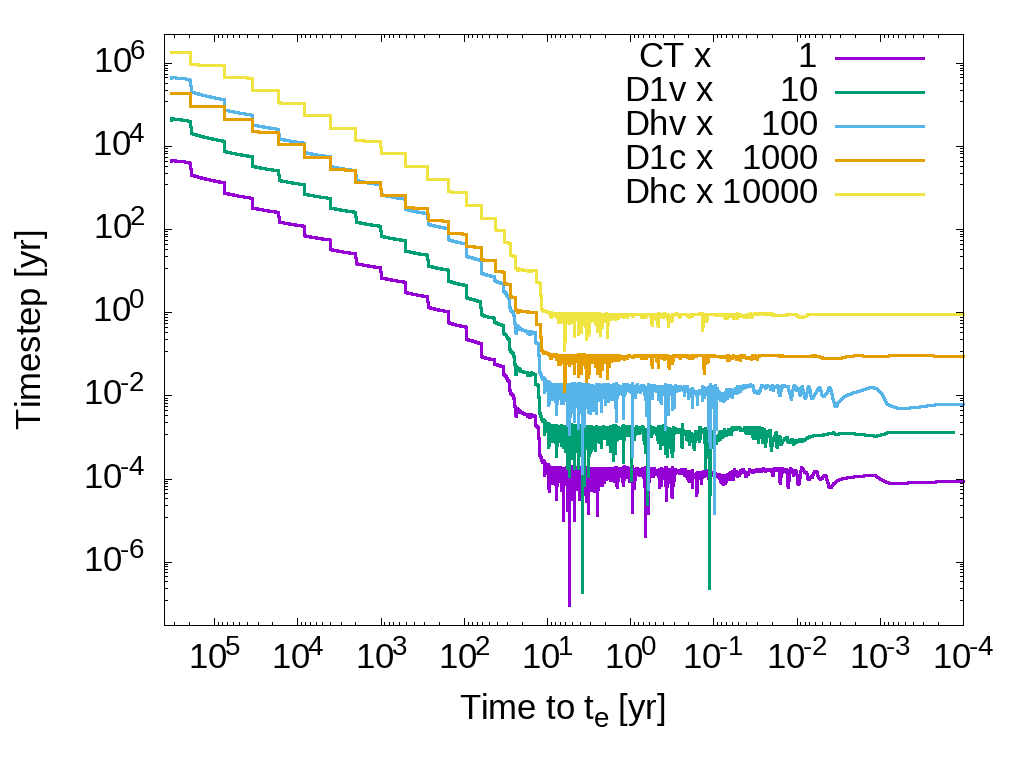}
\end{center}
\caption{Time evolution of the timesteps as a function of the time to $t_e$ (or the lookback time). The lines are plotted with offsets. \label{fig:collapsedt}}
\end{figure}

\subsubsection{Model Setup}
Similar to \citet{athenamg}, we set up a critical Bonnor-Ebert like sphere \citep{bonnor,ebert} with temperature $T=200\,{\rm K}$ and make it unstable by multiplying a density enhancement of $f=6.1$. The resulting mass, central density and radius of the cloud are $M=4,830 \, M_{\odot}$, $\rho_0=6.13\times 10^{-20}\, {\rm g\, cm^{-3}}$ and $R=3.99\times 10^5\, {\rm au}$, respectively. The cloud has a solid-body rotation of $\Omega=1.18\times 10^{-14}\, {\rm s^{-1}}$, corresponding to $\Omega t_{\rm ff}=0.1$ where $t_{\rm ff}$ is the initial free fall time at the center. To model the magnetic field amplification from a very weak seed field, we introduce a uniform magnetic field aligned to the rotation axis $B_z=10^{-20} \,{\rm Gauss}$.

The simulation domain is $[-7.43\times 10^5\,{\rm au}:7.43\times 10^5\,{\rm au}]^3$, and the root grid is resolved with $128^3$ cells. We use AMR with \texttt{MeshBlocks} consisting of $16^3$ cells to resolve the local Jeans length at least with 32 cells. The boundary conditions are set to model an isolated cloud confined by the ambient gas, and are the same as in \citet{athenamg}.

We assume the ideal equation of state of gas with a constant mean molecular weight of $2.3$ for simplicity. While \citet{hirano22} used a realistic table to model the thermal evolution of the gas, we adopt the barotropic approximation for simplicity connecting three regimes: (1) initial temperature of $T=200\,{\rm K}$, (2) runaway collapse with the adiabatic index $\Gamma_{\rm l}=1.1$, and (3) quasi-hydrostatic core formation\footnote{Such core formation should occur at a higher density in the actual Pop-III star formation. This is introduced to prevent further collapse in order to continue calculations at reasonable costs.} with the adiabatic index $\Gamma_{\rm h}=2.0$. 
\begin{equation}
p=\left\{
\begin{array}{ll}
\rho c_{\rm s,0}^2 & (T \le 200\,{\rm K})\\
\rho c_{\rm s,0}^2
\left[
\left(\frac{\rho}{\rho_0}\right)^{2(\Gamma_{\rm l}-1)}
+\left(\frac{\rho}{\rho_{\rm crit}}\right)^{2(\Gamma_{\rm h}-1)}
\right] & (T > 200\,{\rm K})
\end{array}
\right.,
\end{equation}
where $c_{\rm s,0}=0.85\, {\rm km\, s^{-1}}$ is the sound speed at $T=200\,{\rm K}$, and $\rho_{\rm crit}=3.8\times 10^{-8}\,{\rm g\,cm^{-3}}$ is the critical density where the gas becomes stiff. These models are similar to but not exactly the same as those in \citet{hirano22}, because we design our models for demonstrating the behaviors of the different schemes in a simplified configuration.

In the D1v and D1c models, $L=1$ in our unit system corresponds to $6.18\times 10^4\, {\rm au}$ in physical units. This is comparable to a fraction of the initial core radius, and is much larger than cores and disks which form as outcomes of the collapse. On the other hand, the Dhv and Dhc models use $L=h_{\rm min}$, which changes in time depending on the finest AMR level. We adopt a constant transport speed of $c_h=50 c_{\rm s,0}=42.5\,{\rm km\,s^{-1}}$ in D1c and Dhc, which is chosen to be larger than the largest signal speed in the system for the most of the time (but not always; see discussion below) based on the results of CT.

\subsubsection{Results}
We present the results at $t=t_e$ which is about 13 years after the formation of the quasi-hydrostatic core in Figure~\ref{fig:collapse}. At this epoch, the maximum density reaches $\rho\simeq 7.3\times 10^{-6}\,{\rm g\, cm^{-3}}$. Two dense cores with circumstellar disks are already formed, and two more fragments are forming. All the models exhibit consistent dynamical evolutions because magnetic fields are still weak and dynamically insignificant. However, the magnetic field distributions are substantially different between the models. 

Overall, CT (b) and D1c (d) are largely consistent. The magnetic fields are not drastically amplified in these models. The magnetic field distributions correlate with the gas density, indicating that the fields are mainly amplified by the contraction, although additional amplification by the differential rotation plays a significant role as well (see below). The maximum field strength is slightly smaller in D1c, indicating that the divergence cleaning scheme is more diffusive compared to CT. On the other hand, there are noticeable differences on small scales (Figure~\ref{fig:collapsez}). In particular, the magnetic fields around the circumstellar disks are stronger and pointing in the opposite direction in D1c compared to CT. We attribute this difference to the ``leakage" effect of the localized magnetic fields in the divergence cleaning method as discussed in Section~\ref{sec:khloc}.

In contrast, the magnetic fields in D1v (c) are completely different. The maximum field strength is five orders of magnitude larger than in CT and D1c, and the ripple-like patterns are generated on the disk scale and propagate toward the outer envelope. The ripples become wider as they propagate outward, which is correlated with the resolution. The behavior of these ripples is similar to those in the KHI simulations in Section~\ref{sec:khdt}.

The difference between the Dhv (e) and Dhc (f) models is not as drastic as that between the models with $L=1$. However, the fields in both of these models are significantly more amplified by 3--4 orders of magnitude compared to CT and D1c. While the magnetic fields are still relatively localized within the disk structures, some weak ``leakage" fields are visible outside.

Let us estimate the magnetic field amplification solely by gravitational collapse. Assuming quasi-spherical collapse of a uniform cloud, and with $r$ being the cloud radius, the density increases as $\rho \propto r^{-3}$ but the magnetic field scales as $B \propto r^{-2}$. Therefore, a power-law scaling of $B\propto \rho^{2/3}$ is expected. In the present situation, as the maximum gas density increased by more than 14 orders of magnitude, the magnetic fields should be amplified by a few $\times 10^9$ solely by gravitational collapse. The magnetic fields in the CT and D1c models are almost two orders of magnitude larger than this estimate, indicating that the magnetic fields are also amplified by the differential rotation in the small-scale disks. The correlation between the amplified magnetic fields and the rotating disk structure also supports this picture. However, the magnetic fields do not exhibit any steep amplification nor propagation toward the outer envelope shown in \citet{hirano22}. 

\subsubsection{Timestep Evolution}
As seen in Section~\ref{sec:khdt}, the timestep change affects the behavior of the divergence cleaning with variable transport speed. We plot the evolution of the timesteps actually used in the simulations in Figure~\ref{fig:collapsedt}. In the early phase, the timestep decreases gradually as the gravitational collapse proceeds. When new AMR levels are generated, the timestep decreases by a factor of 2. After the formation of the quasi-hydrostatic cores, no more AMR levels are created and the timesteps remain more or less constant. In all the models, the timesteps occasionally get short for an instance (seen as downward spikes). In the present simulations, these are caused by transient numerical errors. For example, numerical fluxes near a strong discontinuity can be inaccurate and produce a cell with very low (or even negative) density or pressure. This kind of numerical error is often observed in simulations of practical applications. Usually such errors are transient and local, and can disappear after a short time (or are treated by some ad hoc methods like density/pressure floors). While gradual timestep changes are less problematic, these abrupt and drastic timestep changes produce artificial magnetic fields in the divergence cleaning methods with the variable transport speed. In practical applications, additional physical processes such as radiation cooling and non-ideal MHD effects may cause similar timestep changes.

The divergence cleaning methods with constant transport speed (D1c and Dhc) do not suffer from this issue. However, these models are more expensive because of the shorter timesteps. In addition, it should be noted that these models still experience short timesteps in spite of the large transport speed, which means that the maximum signal speed exceeds the transport speed for short intervals. We do not find any significant development of numerical instability due to the insufficient transport speed, probably because they are only for short duration and localized, but theoretically, we should use even larger transport speed so that the transport speed remains always the fastest in the system for stability.

\subsection{Present-Day Star and Disk Formation}\label{sec:presentday}
\begin{figure*}[t!]
\begin{center}
\includegraphics[bb=0 0 1915.767409 1665.232439,width=\hsize]{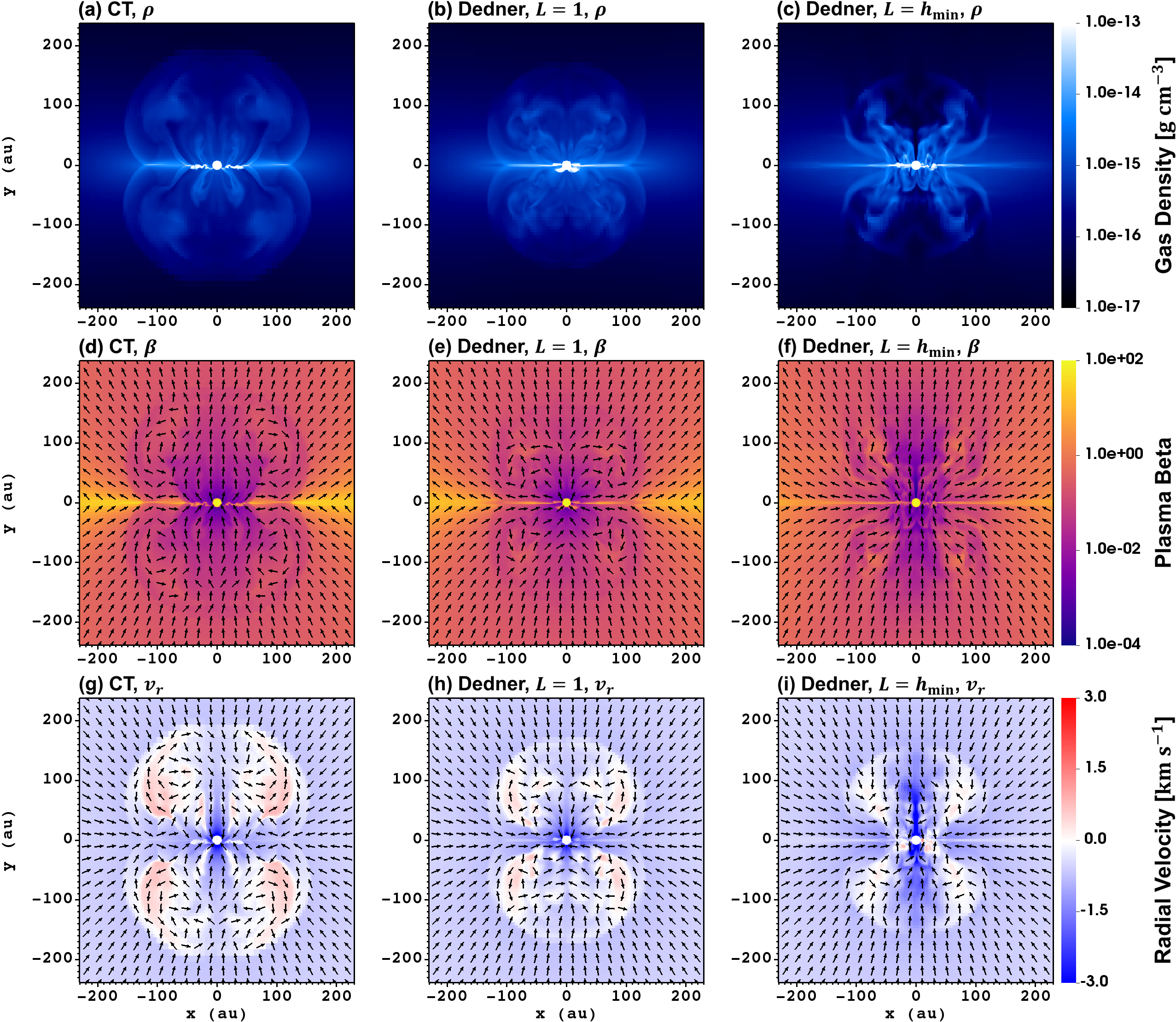}
\end{center}
\caption{Results of the simulations of the collapsing clouds in the present-day environment. From left to right: CT, D1v, and Dhv. From top to bottom: the gas density, plasma beta with magnetic field directions, and radial velocity with flow directions. \label{fig:presentday}}
\end{figure*}

We present another set of test calculations of collapsing clouds in the context of the present-day star formation. This problem has been extensively studied by various groups using different methods, including CT, Dedner's cleaning and the Powell method \citep[e.g.][]{tomida13,tomida15,tsukamoto15,masson16,wurster16,vaytet18,lam19,xu21,tu24,mauxion24,mayer25}, and most of them agree at least qualitatively.

\subsubsection{Model Setup}
We perform the test calculation presented in \citet{athenamg}. This is similar to the previous test, but for a low-mass cloud with a stronger magnetic field. Here we summarize the key model setups, and we refer readers to \citet{athenamg} for details. The initial condition is a $1M_\odot$ Bonnor-Ebert like sphere \citep{bonnor,ebert} with $T=10\,{\rm K}$. Solid-body rotation with $\Omega=5.55\times 10^{-14}\, {\rm s^{-1}}$ and uniform magnetic field $B_z = 27.4\,{\rm \mu G}$ both aligned to the $z-$axis are imposed. We adopt the barotropic relation of
\begin{eqnarray}
P=\rho c_{\rm s}^2\left[1+\left(\frac{\rho}{\rho_{\rm crit}}\right)^{2(\Gamma-1)}\right]^{1/2},
\end{eqnarray}
with $c_{\rm s}=0.19\, {\rm km\, s^{-1}}$, $\rho_{\rm crit}=10^{-13}\,{\rm g\,cm^{-3}}$, and $\Gamma=5/3$. We use AMR to resolve the local Jeans length at least with 32 cells. We compare CT, D1v and Dhv, with $L=1302.7\,{\rm au}$ for D1v.

We use the ideal MHD approximation, although non-ideal MHD effects play significant roles in actual star forming clouds. It should be noted that we perform these calculations purely for demonstrative purposes. In these calculations, very thin current sheets form on the mid-plane as a consequence of the gravitational collapse. These current sheets are so thin that they are practically magnetic discontinuities at the given resolution, even with AMR. In such a situation, numerical solutions are sensitive to the details of the solvers, and do not consistently converge. In other words, this means that the differences between the schemes can appear prominently. Being aware of this limitation, we limit ourselves to qualitative comparison in this work. We emphasize that we would need to consider physical resistivities due to the non-ideal MHD effects and resolve the current sheets in order to obtain consistent and well-converged solutions for scientific studies.

\subsubsection{Results}
Figure~\ref{fig:presentday} shows the vertical cross sections of the gas density, plasma beta with magnetic field directions, and radial velocity with flow directions when the central temperature reaches $T_c \sim 1,000\,{\rm K}$. Adiabatic cores formed \citep[the first Larson cores,][]{larson69} at the center of the clouds, which are almost spherical as a result of efficient angular momentum transport by magnetic fields. Pseudo-disks (disk-like structures but not supported by rotation) form around the first cores, and the inner pseudo-disks are strongly perturbed by the magnetic interchange instability \citep{spruit95}. From the pseudo-disks, slow bipolar outflows with wide opening angles are launched. The properties of the first cores such as the radius and mass are consistent among the three models. On the other hand, the structures in the pseudo-disk and outflows exhibit some differences as expected. 

First, let us compare CT (left) and D1v (middle). The outflow in CT is faster ($v_r\sim 0.9\,{\rm km\,s^{-1}}$ vs $0.6\,{\rm km\,s^{-1}}$) and propagates further ($z\sim 180\, {\rm au}$ vs $130\,{\rm au}$) than in D1v. Also, while the outflow in CT maintains more laminar conical shapes, D1v produces more complex morphology inside the outflow. Inside the outflows, a larger volume is more strongly magnetized (lower plasma beta) in CT compared to D1v, which can be the origin of the difference in the outflow velocities. Despite detailed differences arising from grid-scale behavior of the schemes, these schemes produce qualitatively similar results.

On the other hand, we find a more significant difference in the outflow scale between CT (or D1v) and Dhv. The outflow in Dhv is smaller ($z\sim 85\,{\rm au}$) because its velocity is slower ($v_r \sim 0.3\,{\rm km\,s^{-1}}$), and its morphology is more constricted near the midplane. It is also noticeable that low-density and fast-infalling cavities with low plasma beta are opened up along the polar directions in Dhv.

\subsection{Convective Dynamo in Stratified Atmosphere}\label{sec:convection}
Dynamo in stratified convective plasmas is a key process in magnetic field amplification in various astrophysical systems. As we have seen in the KHI problems, magnetic fields in vortices produce divergence errors, suggesting that the divergence cleaning method is less optimal for dynamo simulations. Here, we present results of simulations of a magnetized, gravitationally stratified convective atmosphere.

\subsubsection{Model Setup} \label{subsec:model_setup}
\begin{figure*}
\centering
\includegraphics[bb=0 0 2272.817247 1194.166686, width=\hsize]{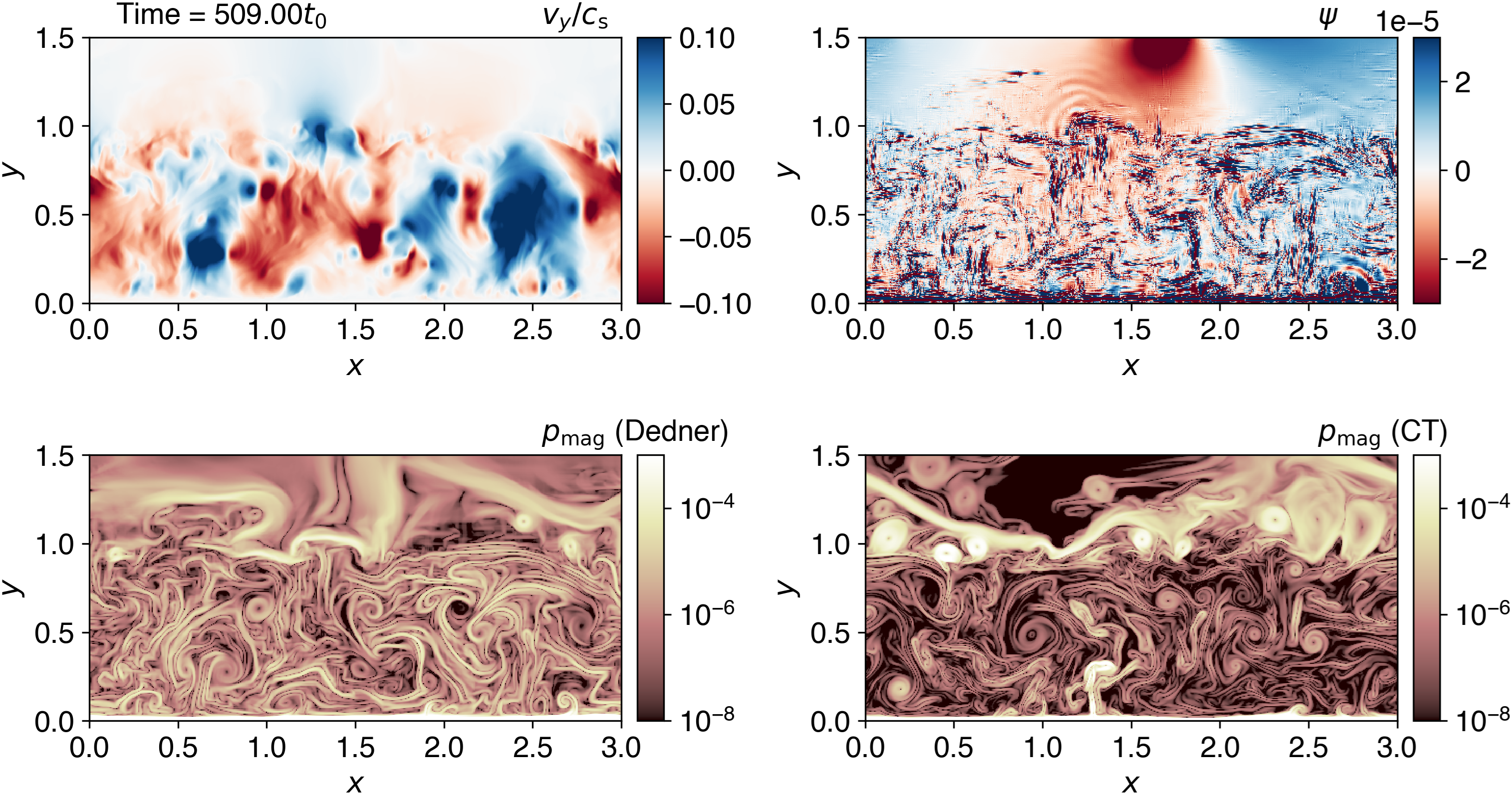}
\caption{From top left to bottom right: $v_y/c_{\rm s}, \psi$, $p_{\rm mag}$ for Dedner's method and $p_{\rm mag}$ for the CT method. The data are taken at $t=509$ in the numerical unit. \label{fig:convection_overview}}
\end{figure*}
\begin{figure}
\centering
\includegraphics[bb=0 0 432.000864 504.001008, width=\columnwidth]{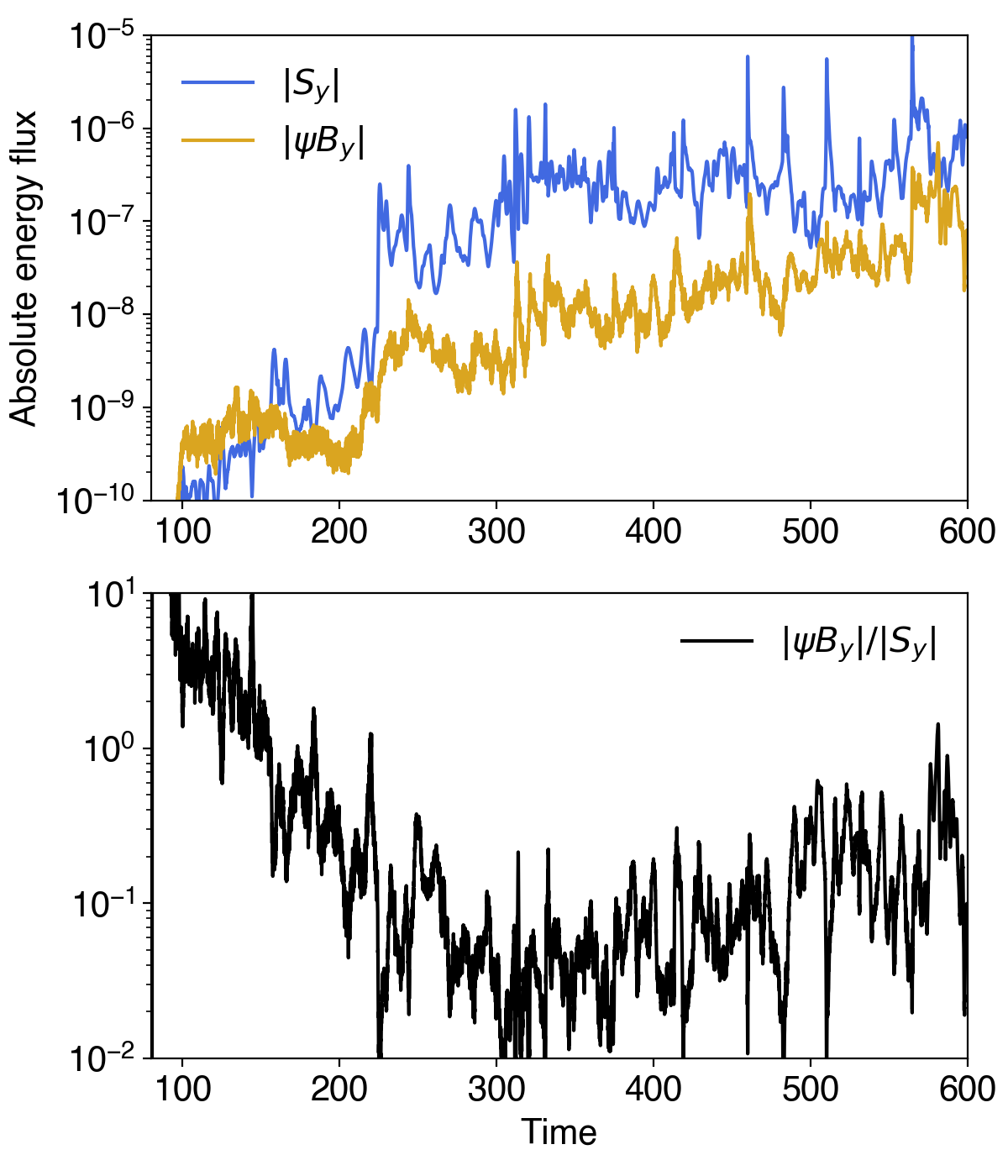}
\caption{Evolution of the energy fluxes measured at the height of $y=1.1$. Top: the absolute values of the vertical component of the Poynting flux (blue) and the $\psi$-related flux (gold). Bottom: the ratio of the two. \label{fig:energy_flux_ratio}}
\end{figure}

\begin{figure}
\centering
\includegraphics[bb=0 0 432.000864 648.001296,width=\columnwidth]{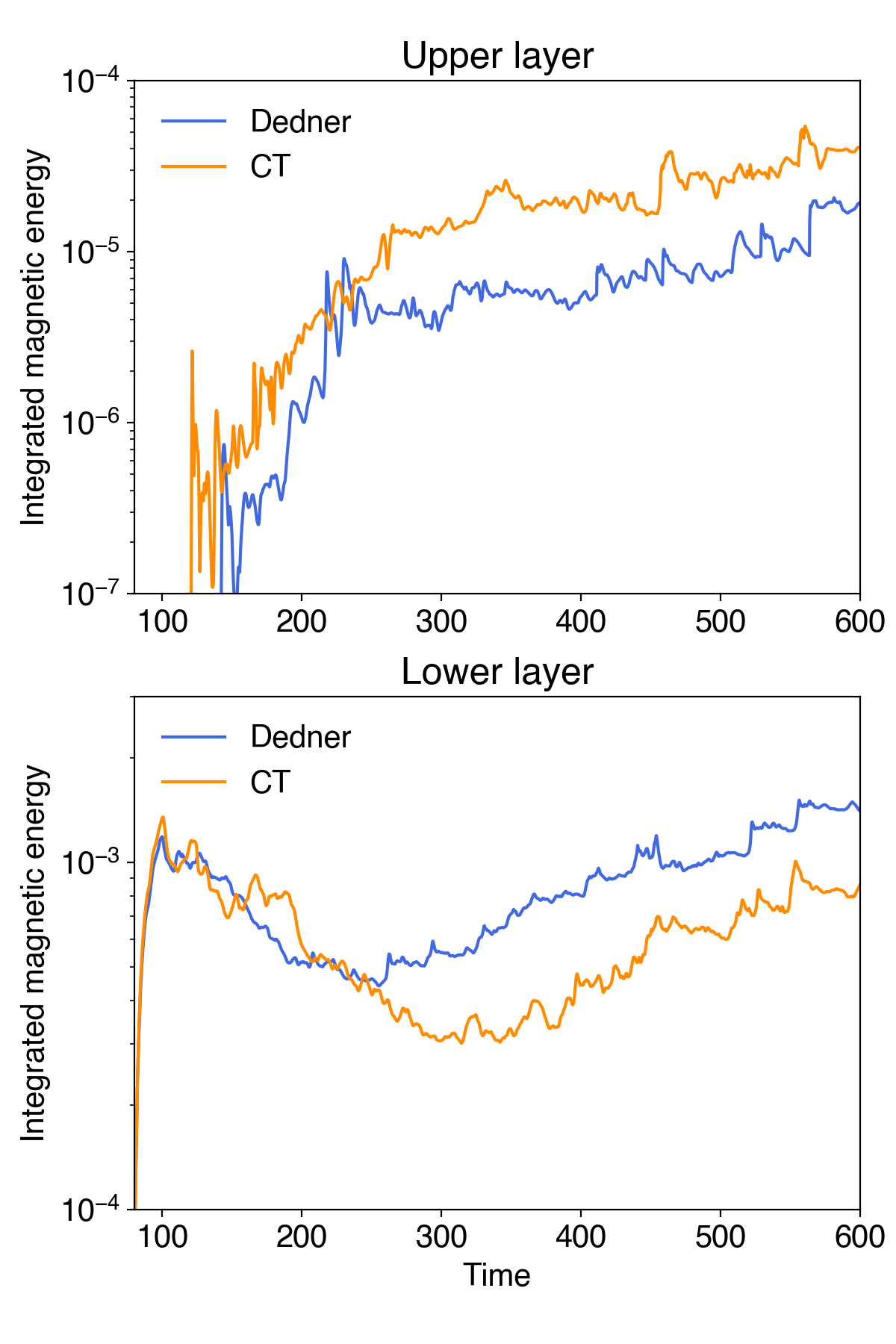}
\caption{Temporal evolution of the magnetic energies. The top and bottom panels show the magnetic energies integrated in $y\ge 1.1$ and $y<1.1$, respectively. The blue and orange lines denote the results for Dedner's method and CT method, respectively. \label{fig:comparison_ME}}
\end{figure}

Our model is a two-dimensional atmosphere which consists of a lower convective layer and an upper convectively stable layer, imitating a stellar surface region of solar-type stars. This setup is motivated to model a convective atmosphere \citep[e.g.,][]{stein98,vogler05}, but is highly simplified because we solely intend to demonstrate the difference between the MHD schemes. The horizontal and vertical coordinates are denoted by $x$ and $y$, respectively, where $0\le x \le 3$ and $0\le y \le 1.5$. The computational domain is resolved with $512 \times 256$ cells.  The equation of state for ideal gas with adiabatic index $\gamma=5/3$ is used. The gas is subject to a uniform downward gravitational acceleration $g_y=-1.4$. We impose periodic boundary conditions in the horizontal direction. At the lower boundary we use a stress-free, hydrostatically extrapolated boundary with reflected vertical velocity. The lower boundary acts as a perfectly conducting boundary for magnetic fields. On the other hand, the upper boundary is a diode boundary: the flow going out of the numerical domain is allowed, while the incoming flow is prohibited. The density and pressure in the ghost cells are computed by hydrostatic extrapolation. The other hydrodynamic quantities and magnetic field components are copied from the last active cells in the calculation domain.

The initial atmosphere is hydrostatic. The prescribed temperature profile, $T_0(y)$, connects an adiabatic lower layer to an isothermal upper layer through a smooth transition,
\begin{equation}
  T_0(y) =
  \left[1-f_{\rm int}(y)\right]T_{\rm ad}(y)
  + f_{\rm int}(y)T_{\rm iso},
\end{equation}
where
\begin{eqnarray}
  f_{\rm int}(y) &=&
  {1\over 2}\left[1+\tanh\left({y-y_{\rm int}\over w_{\rm int}}\right)\right],\\
  T_{\rm ad}(y)&=&T_{\rm bot}
  - {|g_y|(\gamma-1)\over\gamma}(y-y_{\rm bot}).
\end{eqnarray}
We use $T_{\rm bot}=1$, $p_{\rm bot}=1$, $y_{\rm bot}=0$, $y_{\rm int}=1.0$, $w_{\rm int}=0.03$, and $T_{\rm iso}=0.56$. The pressure is obtained from hydrostatic balance. 
The minimum and maximum densities in the numerical domain are approximately 0.067 and 1.0, respectively.
At the beginning of the simulations, we introduce weak random vertical velocity perturbations to the convective layer.

Thermal convection is driven by localized heating near the bottom and Newtonian cooling near the top. We add the volumetric thermal source term to the total energy equation as
\begin{equation}
  Q(y,\rho,T) =
  {F_{\rm heat}\over H_{\rm heat}}
  \exp\left[-{y-y_{\rm bot}\over H_{\rm heat}}\right]
  - {\rho\left(T-T_{\rm cool}\right)\over(\gamma-1)t_{\rm cool}}
  f_{\rm cool}(y),
\end{equation}
with
\begin{equation}
  f_{\rm cool}(y) =
  {1\over2}\left[1+\tanh\left({y-y_{\rm cool}\over w_{\rm cool}}\right)\right],
\end{equation}
where $F_{\rm heat}$ denotes the heat flux, $H_{\rm heat}$ indicates a spatial scale for the heating, $t_{\rm cool}$ is the relaxation timescale for the cooling term, and $y_{\rm bot}, y_{\rm cool}$ and $w_{\rm cool}$ are spatial scales to control the spatial distributions of the heating and cooling terms.
The adopted parameters are $F_{\rm heat}=2\times10^{-4}$, $H_{\rm heat}=0.04$, $t_{\rm cool}=0.05$, $T_{\rm cool}=0.56$, $y_{\rm cool}=y_{\rm int}=1.0$, and $w_{\rm cool}=0.01$.

The magnetic field is initially zero.  After the convection has developed, at $t_{\rm mag}=80$, we insert a horizontal magnetic sheet centered at $y_{\rm mag}=0.20$ with full thickness $w_{\rm mag}=0.10$.  Its profile is
\begin{equation}
  B_x(y) =
  \begin{cases}
  B_0 {1\over2}\left[1+\cos\left({2\pi(y-y_{\rm mag})\over w_{\rm mag}}\right)\right],
  & |y-y_{\rm mag}| < w_{\rm mag}/2,\\
  0, & \mathrm{otherwise},
  \end{cases}
\end{equation}
where $B_0$ is chosen from $\beta_{\rm mag}=2p_0(y_{\rm mag})/B_0^2=1000$.

For this test problem, we adopt the third-order Runge-Kutta time integrator and third-order piecewise parabolic method (PPM) using primitive variables for spatial reconstruction to better resolve turbulent flows. For the calculation with Dedner's method, we use the D1v method with variable $c_h$ and $L=1$.

\subsubsection{Results}

Figure~\ref{fig:convection_overview} presents an overview of the simulated atmosphere. The top left panel (Mach number) demonstrates that convection mainly occurs below $y=1$. The top right panel ($\psi$) indicates that divergence errors are widely generated in the convective layer, as expected. It should be noted that the divergence error produced in the convective layer is propagating toward the upper, convectively stable layer. The panel shows examples of propagating wave fronts around $(x,y)=(1.2,1.2)$. Therefore, this calculation indicates that the magnetic fields are artificially produced in the convectively stable layer due to the divergence error propagation.

To quantitatively evaluate the impact of the divergence error propagation, we measure the energy fluxes associated with magnetic fields at the height of $y=1.1$. The top panel of Figure~\ref{fig:energy_flux_ratio} compares the Poynting flux and the $\psi$-related energy flux (see Appendix~\ref{sec:eglm}), and the bottom panel displays the ratio of the two. It is shown that the contribution of the divergence error does not diminish and remains on the order of $\mathcal{O}(0.01-1)$, which indicates that the divergence error generated in the convective layer has significant impact on the magnetic field evolution in the upper, convectively stable layer.

The bottom panels of Figure~\ref{fig:convection_overview} compare the magnetic pressure of the divergence cleaning and CT methods. The magnetic pressure in the divergence cleaning model is more diffusive than that in the CT model. The magnetic fields spread out smoothly in the upper layer in the divergence cleaning, while the magnetic fields are more localized in CT. In the convective layer, Dedner's method produces more strongly magnetized flows on average than the CT method. Figure~\ref{fig:comparison_ME} highlights the difference between the results of the two schemes. We integrate the magnetic energy in both the upper ($y\ge 1.1$) and lower ($y<1.1$) atmospheres. In the upper layer, the result of the CT method shows a larger magnetic energy, even though the magnetic fields spread out more widely in Dedner's method. In the lower convective layer, on the other hand, the divergence cleaning method yields a larger magnetic energy.

We interpret the difference between the two schemes as follows. In the upper layer, the CT method produces multiple distinct magnetic islands, while Dedner's method only displays a few diffusive ones. This result indicates that CT is less diffusive. We also note that while very weakly magnetized regions are formed in the CT method, such regions are not present in Dedner's method partly because the divergence error from the lower atmosphere propagates everywhere in the upper layer and produces magnetic fields there. Regarding the behavior in the lower convective layer, we argue that the non-local properties of the divergence cleaning method are crucial. The convective layer is full of vortices, and the divergence error generated in a vortex propagates to other vortices and unphysically generates magnetic fields there. As a result, the unphysically produced magnetic fields are further amplified by vortical motions (dynamo), leading to the unphysical formation of a strongly magnetized convective layer.

\section{Discussions}
\subsection{Parameters of the Divergence Cleaning Scheme}
We find that the quality of solutions largely depends on the choice of the parameters $L$ (or $C_r$) and $c_h$ in the divergence cleaning methods, and therefore these values must be selected carefully.

\subsubsection{Characteristic Scale $L$}
From the numerical experiments above, we find that solutions with large $L$ tend to be more consistent with CT, while the use of small (grid scale) $L$ produces unphysical structures. Therefore, we recommend choosing $L$ to be larger than the scale of interest. In addition, for consistency with the derivation, we recommend using constant $L$ both in time and space.

\subsubsection{Transport Speed $c_h$}
Our results clearly indicate that we should use a constant transport speed $c_h$. The fact that the system remains stable even with $c_h\sim 1$ is interesting and deserves discussion.  We compare the transport speeds with characteristic speeds in the system. Because this is a subsonic problem, the sound speed remains almost constant $c_s=\sqrt{\gamma p/\rho}\sim 4.08$. The maximum fluid speed at $t=3.0$ is $v_{\rm max}\sim 1.38$, and the maximum Alfv\'{e}n speed is $v_{a,{\rm max}}\sim 0.36$ in the high resolution CT model. While \citet{arepo} argued that $c_h$ should be larger than the fastest signal speed in the system and proposed to use the largest speed of fast magnetoacoustic waves, our results imply that we can use a smaller value for $c_h$. 

In order to investigate the stability criterion, we have performed additional numerical experiments with increased Alfv\'{e}n speed and increased flow velocity for a longer period. Interestingly, the system remains stable even when the Alfv\'{e}n speed exceeds $c_h$. However, the system gets unstable with the flow velocity exceeding $c_h$. Thus, our experiments suggest that the transport speed $c_h$ should be larger than the maximum flow speed, but it is not directly relevant to the other speeds like the sound speed or Alfv\'{e}n speed. However, we still recommend conservatively choosing $c_h$ larger than the maximum signal (total) speed, i.e., $c_h=\max\left(\sqrt{v^2+c_f^2}\right)$ because the thermal and/or magnetic energy can be converted into the kinetic energy and the flow can be accelerated to a comparable speed in actual simulations\footnote{Note that this can depend on the base schemes. For example, \citet{arepo} uses $c_h=\max(c_f)$ in AREPO, which is a Lagrangian moving mesh code. Also, the behavior may change if we use the Galilean-invariant version of the mixed cleaning scheme \citep[eq.38 of][]{dedner}}.

Use of such a large transport speed is computationally expensive because it requires significantly smaller timesteps. Also, it is not trivial to predict the fastest speed throughout a simulation without knowing the system's behavior in advance. Our results imply that gradual change in the timestep can be acceptable if it is sufficiently slow (i.e., the non-commutative effect is negligible or processed quickly by the damping term), but it is not always possible because a short timestep can be suddenly required by numerical glitches or other physical processes\footnote{Similarly, spatially non-uniform transport speed may work if there is no steep gradient. However, it is difficult to control because discontinuities commonly appear in astrophysical simulations.}. If it is really necessary to change the transport speed (or any other parameter) during a simulation, we propose the following ``refresh" procedure: apply the projection method \citep{brackbill80} to eliminate the divergence error and reset $\psi$ to 0 everywhere in the computational domain before changing the transport speed. This should work because the transport speed can be changed without any problem if there is no divergence error in the system, and the additional cost should be acceptable if it is applied only occasionally.

\subsection{Origin of the Spurious Behaviors}
\subsubsection{Leakage of Localized Magnetic Fields}
Through the numerical experiments, we have demonstrated that the divergence cleaning scheme tends to be diffusive and can cause visibly noticeable errors. In particular, we have shown that strongly localized fields still leak into weakly magnetized regions even with the optimal choices of these parameters. This is a natural consequence of the scheme that the divergence error is isotropically transported while being damped. As demonstrated above, use of the stronger damping factor (smaller $L$) can reduce the affected region, but artificial structures can appear if the damping is too strong\footnote{In the limit of $L\rightarrow 0$, $\psi$ remains almost zero as it is immediately damped, and the scheme cannot clean the divergence error effectively because the $\nabla \phi$ term in eq.(\ref{eq:ind2}) vanishes.}. Therefore, while there is room for optimizing the parameters, it is not a fundamental solution to this issue.

\subsubsection{Spurious Field Generation with Timestep Change}
Mathematically, the spurious behavior related to the sudden timestep change is caused by the use of the variable transport speed $c_h$ because it is inconsistent with the derivation. When $c_h$ changes drastically, the divergence cleaning variable $\psi$ does not follow the same equation as $\nabla\cdot\mathbf{B}$ anymore, and non-commutative terms produce the artificial fields.

Let us discuss this behavior more intuitively here. First, we explain the behavior of the $L=1$ model in the KHI test with localized magnetic fields (D1v, Panels (a1) -- (a6) in Figure~\ref{fig:kh_dt1}). The divergence cleaning scheme works reasonably well while the timestep changes only slowly. The amplitude of the divergence cleaning variable is more or less balanced in this situation and behaves roughly $\psi \propto c_h B \propto B \Delta t^{-1}$ (Figures~\ref{fig:kh_dt1} (a1) and \ref{fig:psimax}, see also Figure~\ref{fig:kh_ch}). When the timestep suddenly decreases, the transport speed increases in response, and $\psi$ is quickly amplified until it reaches the new balanced state because the flux $F_\psi$ (eqs.(\ref{eq:bxm}) and (\ref{eq:psiflux})) is enhanced (a2). With relatively large $L$, the damping by the source term is slow and the amplified $\psi$ propagates isotropically at the fast transport speed $c_h$ (a3). When the timestep is suddenly recovered, the amplified $\psi$ is quickly transformed to the magnetic field by the flux $F_{B_x}$ (eqs.(\ref{eq:psim}) and (\ref{eq:bxflux})) (a4). Then, the artificially produced magnetic fields propagate and disrupt the field structures (a5 -- a6). The same mechanism produces the spurious magnetic fields in the collapse simulation in Section~\ref{sec:collapse} (Panel (c) in Figure~\ref{fig:collapse}).

On the other hand, the model with $L=h_{\rm min}$ (Dhv, Panels (b1) -- (b6) in Figure~\ref{fig:kh_dt1}) exhibits the same amplification of $\psi$ with the sudden timestep change but it does not produce significant artificial ripple patterns (b2). This is because the strong damping suppresses $\psi$ and the system reaches a new balanced state quickly, preventing the amplified $\psi$ from propagating further (b3 -- b4).

This spurious behavior does not appear with the constant transport speed $c_h$ because the linear operator $\mathcal{D}$ is commutative with the spatial and temporal derivative operators. In this case, the balance between $\psi$ and $B$ is maintained even when the timestep changes drastically, and the amplification mechanism through the fluxes does not work.

In addition, by comparing the first and second terms on the right-hand side in equations (\ref{eq:bxm}) and (\ref{eq:psim}), this amplification is more severe when the magnetic fields are weak. The error produced by the divergence cleaning algorithm is less pronounced as long as $|\mathbf{B}|\gg \frac{\psi}{c_h}$, although quantitatively we need to compare the force by the artificially produced magnetic fields and the other forces. This qualitatively explains the difference between the test cases in Sections~\ref{sec:collapse} and \ref{sec:presentday}. This also implies that the larger transport speed is more robust, although it is computationally more expensive.

\subsection{Implications on Past Works}
Our results indicate that some previous works using the divergence cleaning methods should be carefully reexamined and verified. To be specific, we identify that the divergence cleaning scheme can fail and produce artificial magnetic fields when magnetic fields are strongly localized or the timestep changes suddenly and drastically. 

\subsubsection{Magnetic Field Amplification in Weakly Magnetized Clouds}
One of the notable astrophysical examples of such a situation is star and disk formation in a collapsing cloud. As we have demonstrated in Section~\ref{sec:collapse}, the divergence cleaning schemes with the variable transport speed drastically produce artificial magnetic fields in the early Universe case, particularly with $L=1$. Our results suggest that the extremely rapid amplification of magnetic fields in collapsing clouds reported in \citet{hirano22, machida25} are likely artifacts due to the divergence cleaning scheme. The ripple-like patterns in Figure~2 of \citet{hirano22} resemble those in our Figure~\ref{fig:collapse}, which also support our argument. Similar rapid amplification of magnetic fields is also reported in \citet{latif14} \citep[see also][]{latif23}, who use the Enzo code with Dedner's divergence cleaning method \citep{enzo,wang09}\footnote{Although the values of the divergence cleaning parameters are not shown in their papers, the public version of the Enzo code (version 2.6) adopts the variable transport speed with $L=1$ by default.}. In their results, no prominent ripple-like pattern is visible \citep[Figure~1 of][]{latif14}, possibly because such features are smeared out by turbulence or by the diffusive nature of the HLL solver. Although they attribute the amplification to strong accretion shocks, we note that, from the MHD Rankine-Hugoniot relation, even a strong shock can only amplify the tangential component of the magnetic field by at most the density compression ratio. Therefore, presence of the strong shock alone is insufficient to explain the amplification. In addition, the magnetic fields are amplified also inside the core in their simulation where no shock exists. If the fields inside the core are transported from the shock rather than amplified locally, it requires unphysically fast transport speed greater than $100\,{\rm km\, s^{-1}}$. Furthermore, the amplified magnetic fields in their simulations appear to leak into the upstream of the shock \citep[Figure~3 of][]{latif14}. These behaviors raise the possibility that the results are affected by numerical artifacts due to the divergence cleaning method, at least in part. While there still should be significant amplification by rotating disks in addition to gravitational contraction, the actual amplification should be much more modest compared to these works. Further studies are needed in order to quantify the amplification of magnetic fields and their significance in the early Universe.

A similar situation can occur in many astrophysical systems where initially weak magnetic fields are strongly amplified in rotating accretion flows, including magnetic field amplification in galaxy formation in cosmological contexts, tidal disruption events, mergers of compact objects, among others.

\subsubsection{Star and Disk Formation in Present-Day Environments}
In the present-day cases with stronger magnetization in Section ~\ref{sec:presentday}, the divergence cleaning scheme and CT produce qualitatively similar structures, which is likely because the magnetic fields are stronger and less localized. This is consistent with the fact that many previous works (see the list in Section~\ref{sec:presentday}) yield qualitatively similar results. However, we should not interpret these results as evidence that all the schemes behave consistently when the magnetic fields are strong. In fact, there are substantial differences in the dynamics; for example, the outflow velocities are different by a factor of $\sim 3$. Therefore, we still need to carefully compare and interpret the results of simulations using different schemes.

It should be noted that these differences are likely produced by the different grid-scale behaviors between the schemes. Because very thin current sheets tend to form in the ideal MHD approximation, the results are sensitive to the details of the schemes. In such situations, it is difficult to obtain converged solutions between different schemes. For consistent comparison, we need to introduce viscosity and resistivity so that the minimum scale structures are well resolved with the grid.

\subsubsection{Convective Dynamo in Stellar Atmosphere}
Another example that requires particular attention is models of magnetized, stratified convective layers. As we have demonstrated in Section~\ref{sec:convection}, the non-local properties of Dedner's method significantly affect the magnetic field evolution. The divergence error generated in the convective layer propagates through all regions, producing unphysical magnetic fields even in the upper, non-convective layer. Additionally, we show that the error propagation in the turbulent convective layer can result in the formation of unphysically strong magnetization in the convective flows. These results have implications for modeling of solar/stellar atmospheres, for example. The unphysical energy transport and magnetic field generation in the chromosphere and corona, as well as artificial magnetic dissipation arising from divergence errors, may cause misleading interpretation of the heating mechanisms. Dynamo calculations using Dedner's method can produce unphysically strong magnetization in convective layers, which may lead to an overestimation of the Maxwell stress that controls the angular momentum transport in rotating systems. We are unsure if such unphysical behaviors diminish when we increase the numerical resolution, as the non-local properties of the scheme remain regardless of the resolution. \\

To draw more reliable conclusions, we should carefully reexamine and verify these results, preferably by testing them against independent methods such as CT.

\section{Summary}
We have implemented Dedner's divergence cleaning method in Athena++ and compared it with CT. Through numerical experiments, we have investigated the optimal choice of the parameters in the divergence cleaning scheme, and have identified certain situations in which the divergence cleaning methods can fail. We propose a few improvements to the scheme to increase its robustness and consistency, but these are not sufficient to completely suppress the spurious behaviors. We summarize our findings as follows:

\begin{itemize}
\item The parameters must be carefully chosen for the divergence cleaning scheme. Both the characteristic scale $L$ and transport speed $c_h$ should be temporally constant and spatially uniform. We recommend using a characteristic scale $L$ larger than the scale of interest, and $c_h$ larger than the maximum signal speed that can occur during the whole simulation, although they require knowledge about the system in advance and can be computationally expensive. The conventional implementation with variable transport speed is strongly discouraged.

\item The solutions with CT and the divergence cleaning method with large $L$ and $c_h$ can be consistent if the magnetization is not strongly localized, with CT being slightly less diffusive for the same resolution and same spatial reconstruction.

\item With variable transport speed, artificial magnetic fields are produced when the timestep changes drastically. Such fields do not satisfy the divergence-free constraint, and exhibit characteristic ripple-like patterns. While we do not recommend it, one should carefully watch for such structures if the divergence cleaning method with variable $c_h$ has to be used.

\item With strongly localized magnetic fields, the divergence error produced in the strongly magnetized region leaks into the weakly magnetized region and produces artificial structures in the divergence cleaning scheme even with a constant transport speed and optimal parameters. This issue persists even with higher resolution. In such a situation, the divergence cleaning scheme must be used with extreme caution.

\item Some results showing strong magnetic field amplification using Dedner's scheme in previous works must be carefully reviewed and reexamined. Further studies with robust numerical methods are needed.
\end{itemize}

In conclusion, CT is less arbitrary, more robust and less prone to spurious behaviors in general. While we do not argue that divergence cleaning methods should be completely abandoned, they should be used with caution. We recommend fixing all the parameters $L, c_p, c_h$ throughout a simulation, with $L$ larger than the scale of interest and $c_h$ faster than the fastest possible speed in the system. In addition, we recommend including the EGLM source term in the energy equation to consistently calculate the contribution from the divergence cleaning variable (see Appendix~\ref{sec:eglm}). However, we would like to emphasize that the divergence cleaning methods can fail even with the optimal parameters, and simulation results obtained with the divergence cleaning methods must be carefully verified based on both physics and numerics.

\begin{acknowledgments}
We thank Tomoaki Matsumoto, Masahiro N. Machida and Yuki Kudoh for fruitful discussions. We also thank Daniel Price for providing a note on the derivation of the energy equation including the divergence cleaning variable. We are also grateful to the anonymous referee for valuable comments and suggestions that helped improve the manuscript. KT acknowledges the visiting professor program of École Normale Supérieure de Lyon, where most of this paper was written. This work is supported by Japan Society for the Promotion of Science (JSPS) KAKENHI Grant Numbers JP21H04487, JP22KK0043 (KT, KI, ST), JP26K00769 (ST), and KES is supported by JSPS Overseas Research Fellowship. KT also acknowledges the 2024 Inamori Research Grant from the Inamori Foundation. This work is also supported by the Ministry of Education, Culture, Sports, Science and Technology (MEXT) of Japan as ``Program for Promoting Researches on the Supercomputer Fugaku" (Structure and Evolution of the Universe Unraveled by Fusion of Simulation and AI; Grant Number JPMXP1020230406). Numerical computations were in part carried out on HPE Cray XD2000 supercomputer at Center for Computational Astrophysics, National Astronomical Observatory of Japan. The original title of this paper was ``Dr. Staggeredgrid, Or: How I Learned to Stop Worrying and Love the Constrained Transport." 
\end{acknowledgments}

\software{Athena++ \citep{athenazenodo}}

\appendix
\section{Powell's 8-wave Method} \label{sec:powell}

Powell's 8-wave method \citep{powell94,powell} is an alternative method to deal with $\nabla\cdot\mathbf{B}$, and is supported in some public codes \citep[e.g.][]{weinberger20}. In this scheme, additional source terms proportional to $\nabla\cdot\mathbf{B}$ are added on the right-hand side of the MHD equations. While this scheme is often referred to as Powell's cleaning, it is actually not ``cleaning" as it does not explicitly eliminate or damp the divergence error. It tends to be more robust than doing nothing, but it is not sufficient to stabilize the system in some applications. See \citet{mocz14,arepoct,gizmo} for example.

\begin{figure}[tb]
\includegraphics[bb=0 0 442.561774 420.058633,width=\columnwidth]{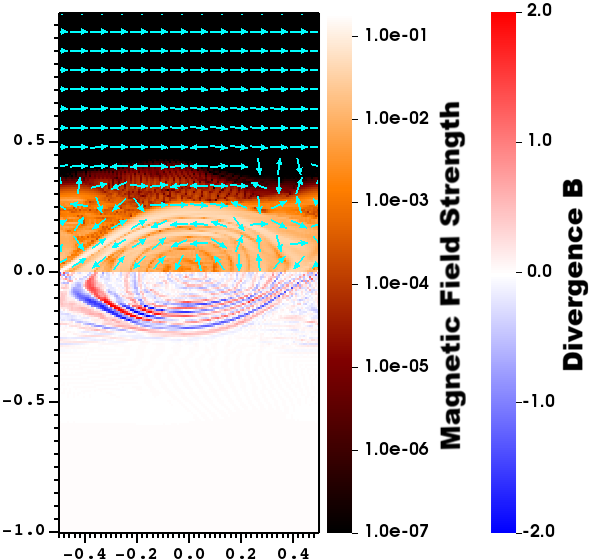}
\caption{Magnetic field strength (upper half) and $\nabla\cdot\mathbf{B}$ at $t=3.0$ in the KHI test with localized magnetic fields with Powell's method. Magnetic field directions are overplotted with cyan arrows in the top half. \label{fig:kh_powell}}
\end{figure}

To show the behavior of this scheme, we implement it in Athena++. All the source terms are discretized using cell-centered differences and integrated explicitly. Figure~\ref{fig:kh_powell} shows the result of the KHI test with localized magnetic fields presented in Section~\ref{sec:khloc}. The amplitude of the divergence error increases with time as expected, but the scheme does not crash (i.e., does not cause a catastrophic explosion or negative density/pressure) even with the substantial divergence error. The magnetic fields show scaly patterns which are clearly artificial. While the artificially produced magnetic fields do not propagate further like in Dedner's method, the fields produced in the vortex still diffuse out to the weakly magnetized region. Thus, this method is also suboptimal compared to CT.

\section{GLM vs EGLM} \label{sec:eglm}
In the original paper of \citet{dedner}, they propose an alternative formula called Extended GLM (EGLM). While some comparisons are presented, the advantage of the scheme is not very clear in the original paper. Here we briefly describe the EGLM method and discuss its implications.

\subsection{EGLM equations}
In EGLM, additional source terms are added in the momentum equation and energy equation.
\begin{eqnarray}
\frac{\partial \rho \mathbf{v}}{\partial t} +\mathbf{\nabla\cdot} \left(\rho\mathbf{vv} - \mathbf{BB} + P^*{\mathbb I}\right)  &=& -(\nabla\cdot\mathbf{B})\mathbf{B},\label{eq:eom2}\\
\frac{\partial E}{\partial t} +\nabla\cdot \bigl[(E + P^*) \mathbf{v} - \mathbf{B} (\mathbf{B \cdot v}) \bigr]&=&-\mathbf{B}\cdot\nabla\psi,\label{eq:eoe2}
\end{eqnarray}
These source terms violate the conservation of the momentum $\rho \mathbf{v}$ and total energy $E$.

The source term in eq.(\ref{eq:eom2}) is introduced by retaining the term which is usually omitted as it should be identically zero if the solenoidal constraint is perfectly satisfied. Conceptually, this is analogous to the Powell's method\footnote{It is not clear why they consider only the momentum equation and do not introduce all the relevant source terms as in the Powell method.}.

On the other hand, the source term in eq.(\ref{eq:eoe2}) has more ``physical" implication. In \citet{dedner}, this term is introduced by considering the contribution of $\psi$ to the magnetic energy. \citet{tricco12} present its meaning more intuitively and implemented it in their SPH code. Because $\psi$ transports the magnetic fields, $\psi$ must carry its own energy as well. To clarify this effect, we transform the right-hand side of the energy equation (\ref{eq:eoe2}) using eq.(\ref{eq:psi}) as follows.
\begin{eqnarray*}
-\mathbf{B}\cdot\nabla\psi&=&-\nabla\cdot(\psi\mathbf{B})+\psi\nabla\cdot\mathbf{B},\\
&=&-\nabla\cdot(\psi\mathbf{B})-\frac{\psi}{c_h^2}\left(\frac{\partial \psi}{\partial t}+\frac{c_h^2}{c_p^2}\psi\right).
\end{eqnarray*}
By defining the total energy including the energy of $\psi$ as $E' = e + \frac{1}{2}\rho v^{2} + \frac{B^{2}}{2}+\frac{\psi^2}{2c_h^2}$, we obtain the new total energy equation:
\begin{eqnarray}
\frac{\partial E'}{\partial t} +\nabla\cdot \bigl[(E + P^*) \mathbf{v} - \mathbf{B} (\mathbf{B \cdot v}) +\psi\mathbf{B}\bigr]&=&-\frac{\psi^2}{c_p^2}.\label{eq:eoe3}
\end{eqnarray}
This means that the energy density of $\psi$ is $e_\psi\equiv\frac{\psi^2}{2c_h^2}$. If there is no source term on the right-hand side corresponding to the damping term, this new total energy $E'$ is conserved. It should be noted that this equation does not hold if $c_h$ is not constant.

This consequence has a significant impact on the behavior of the scheme. While the EGLM energy equation (\ref{eq:eoe2}) is equivalent to considering the total energy conservation including $e_\psi$ (\ref{eq:eoe3}), the original GLM equation misses this contribution and cannot capture the energy transport by $\psi$. Because the thermal energy is calculated by subtracting the kinetic and magnetic energies from the total energy, this error appears in the thermal energy. Therefore, although the total energy (without $e_\psi$) is conserved in the GLM formula, it is not fully consistent with the actual energy transport and can lead to unphysical behaviors. We present a direct comparison of these schemes in Section~\ref{sec:poc}.

\citet{dedner} also proposed Galilean invariant versions (eqs.(38) in their paper). All the source terms same as the Powell's method are introduced on the right-hand side of the MHD equations for these variants, which can improve the robustness of the scheme. Because the transport speed of the divergence error is not isotropic, these schemes may behave differently from the original versions. However, testing these schemes is beyond the scope of this paper.

\subsection{Implementations}
We implement two different approaches of EGLM. The first approach (EGLM-S) is a direct discretization of the source terms in equations (\ref{eq:eom2}) and  (\ref{eq:eoe2}).
\small{\begin{eqnarray}
&&[-(\nabla\cdot\mathbf{B})\mathbf{B}]_{i,j,k}=-\frac{\mathbf{B}_{i,j,k}}{\Delta V_{i,j,k}}\nonumber \\
&\times&[(B_{x,i+1/2,j,k}\Delta S_{x,i+1/2,j,k}-B_{x,i-1/2,j,k}\Delta S_{x,i-1/2,j,k})\nonumber \\
&+&(B_{y,i,j+1/2,k}\Delta S_{y,i,j+1/2,k}-B_{y,i,j-1/2,k}\Delta S_{y,i,j-1/2,k})\nonumber \\
&+&(B_{z,i,j,k+1/2}\Delta S_{z,i,j,k+1/2}-B_{z,i,j,k-1/2}\Delta S_{z,i,j,k-1/2})],
\end{eqnarray}}
\begin{eqnarray}[-\mathbf{B}\cdot\nabla\psi]_{i,j,k}=&-&B_{x,i,j,k}\frac{\psi_{i+1/2,j,k}-\psi_{i-1/2,j,k}}{\Delta x}\nonumber \\
&-&B_{y,i,j,k}\frac{\psi_{i,j+1/2,k}-\psi_{i,j-1/2,k}}{\Delta y}\nonumber \\
&-&B_{z,i,j,k}\frac{\psi_{i,j,k+1/2}-\psi_{i,j,k-1/2}}{\Delta z},
\end{eqnarray}
where $\Delta S$ is the surface area of the cell in each direction and $\Delta V$ is the cell volume. We evaluate magnetic fields and $\psi$ at the cell surfaces using equations (\ref{eq:bxm}) and (\ref{eq:psim}). These source terms are explicitly integrated as part of the main integrator of Athena++ as they should not be very large as long as the scheme is working well. The rest remains exactly the same as in the GLM scheme described in Section~\ref{sec:implementation}.

The second approach (EGLM-C) is based on the conservative form of the total energy equation (\ref{eq:eoe3}) including the contribution from the divergence cleaning variable. In this implementation, we replace the total energy $E$ with $E'$, and modify the primitive-conservative conversion subroutines in Athena++. In order to minimize the change to the existing code, we estimate the new total energy flux as a sum of the total energy flux obtained from the normal HLLD Riemann solver and the $\psi$-energy flux:
\begin{eqnarray*}
\mathbf{F}_{E'}&=&\mathbf{F}_{E}+\mathbf{F}_{e_\psi},\\
\mathbf{F}_{e_\psi}&\equiv&\psi\mathbf{B},
\end{eqnarray*}
and discretize it as 
\begin{eqnarray*}
\mathbf{F}_{e_\psi,x,i+1/2,j,k}=\psi_{i+1/2,j,k} B_{x,i+1/2,j,k},
\end{eqnarray*}
where again $B_x$ and $\psi$ at the cell surfaces are calculated using eqs.(\ref{eq:bxm}) and (\ref{eq:psim}). The source term on the right-hand side in the total energy equation (\ref{eq:eoe3}) is corresponding to the energy loss by the damping of $\psi$. As the damping term is updated exactly using the operator splitting as in eq.(\ref{eq:damp}), we calculate this source term consistently with the damping as follows, instead of directly discretizing it.
\begin{eqnarray*}
\left[-\Delta t\frac{\psi^2}{c_p^2}\right]_{i,j,k}=e^{n+1}_{\psi,i,j,k}-e^{*}_{\psi,i,j,k}=\frac{(\psi_{i,j,k}^{n+1})^2-(\psi_{i,j,k}^{*})^2}{2c_h^2}.
\end{eqnarray*}
We use the same discretization for the source term in the momentum equation (\ref{eq:eom2}). The other equations remain the same as in the GLM scheme.

\subsection{Proof of Concept Calculations}\label{sec:poc}
\begin{figure*}[t]
\includegraphics[bb=0 0 1447.702075 712.599467,width=\textwidth]{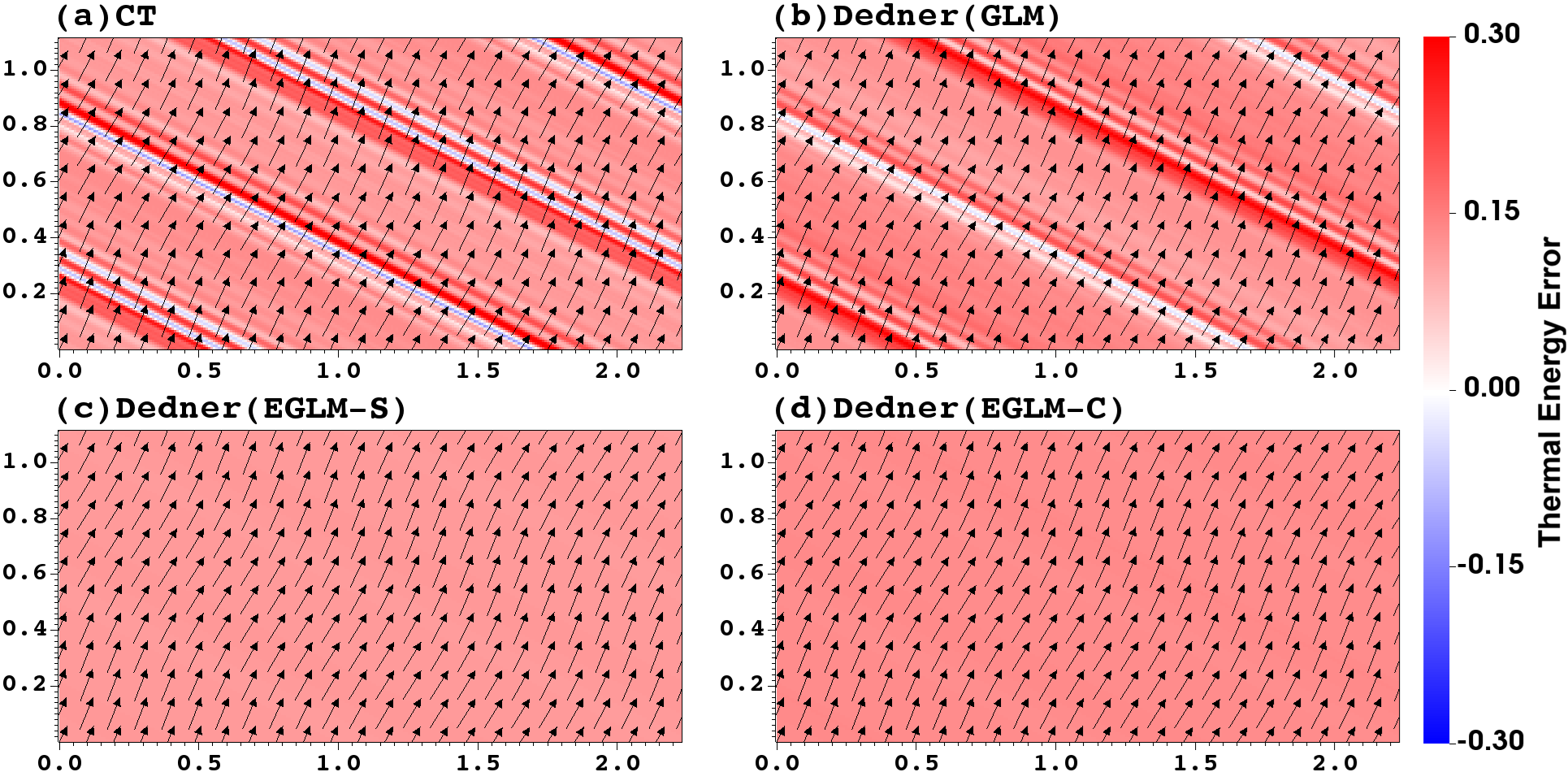}
\caption{Distributions of the thermal energy error with magnetic field directions in the circularly polarized Alfv\'{e}n wave test at $t=5.0$. \label{fig:cpaw1}}
\end{figure*}

\begin{figure}[t]
\includegraphics[bb=0 0 1024.000000 768.000000,width=\columnwidth]{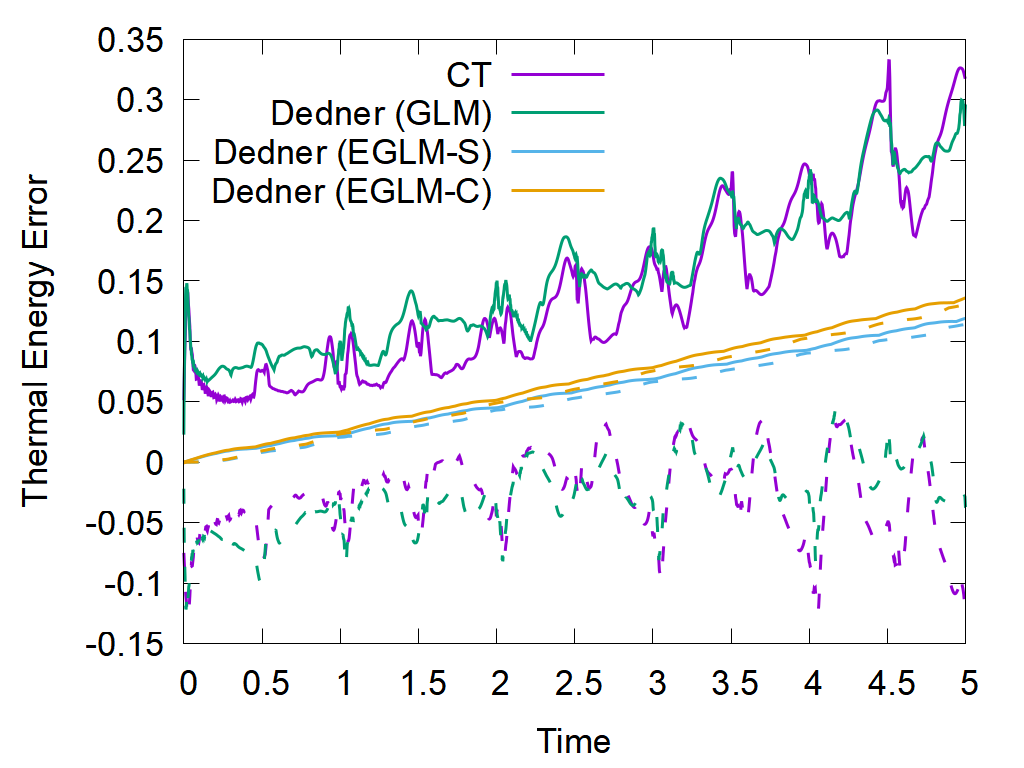}
\caption{The time evolution of the maximum (solid) and minimum (dashed) thermal energy errors in the circularly polarized Alfv\'{e}n wave test. \label{fig:cpaw2}}
\end{figure}

In this section, we demonstrate the difference between the CT, GLM and EGLM schemes. From the discussion above, it is expected that the difference should appear in the thermal energy. We use the circularly polarized Alfv\'{e}n wave in two dimensions as the test problem \citep{toth00}. This problem has an analytic solution, and because the Alfv\'{e}n wave is incompressible, the thermal energy should remain unchanged. However, in practice, the thermal energy should gradually increase because non-zero numerical diffusion dissipates the kinetic and magnetic energies.

We adopt the setup of \citet{gs05}, in which the Alfv\'{e}n wave travels at an angle of $\theta=\tan^{-1}(2)$ with respect to the $x$-axis and has a wavelength of $\lambda=1$. The computational domain spans $[0:\sqrt{5}]\times[0:\sqrt{5}/2]$ and is resolved with $256\times 128$ cells. The initial density and pressure are uniform, and $\rho_0=1$ and $p_0=10^{-4}$. To initialize the velocity and magnetic fields, we use a rotated coordinate system aligned with the traveling direction of the Alfv\'{e}n wave:
\begin{eqnarray*}
x_1 &=& x\cos\theta +y\sin\theta,\\
x_2 &=& -x\sin\theta +y\cos\theta,\\
x_3 &=& z,
\end{eqnarray*}
and in this coordinate system, $\mathbf{v}$ and $\mathbf{B}$ are initialized as
\begin{eqnarray*}
v_1 &=& 0,\\
v_2 &=& v_\perp\sin(2\pi x_1),\\
v_3 &=& v_\perp\cos(2\pi x_1),\\
B_1 &=& B_0,\\
B_2 &=& B_\perp\sin(2\pi x_1),\\
B_3 &=& B_\perp\cos(2\pi x_1),
\end{eqnarray*}
where $v_\perp=0.1$ and $B_\perp=0.1$ set the amplitude of the wave. We adopt $B_0 = 1$, corresponding to the Alfv\'{e}n speed of $v_a=1$ and the plasma beta of $\beta=1.98\times 10^{-4}$. We adopt such a low plasma-beta configuration intentionally in order to elucidate the difference between the schemes in a stringent condition. The boundary conditions are periodic in all the directions, and the adiabatic index is $\gamma=5/3$. For GLM and EGLM, we use $L=1$ and variable transport speed (corresponding to D1v).

We measure the fractional error of the thermal energy
\begin{eqnarray*}
\delta e_{i,j}\equiv \frac{e_{i,j}-e_0}{e_0},
\end{eqnarray*}
where $e_0$ is the initial thermal energy. We present the distributions of $\delta e$ at $t=5.0$ (corresponding to 5 periods) in Figure~\ref{fig:cpaw1} and the time evolution of the maximum and minimum values in Figure~\ref{fig:cpaw2}. As discussed above, $\delta e$ should remain zero in the exact solution. CT and GLM produce considerable error in the thermal energy with stripe patterns, both positive and negative, and its amplitude reaches $\sim30\%$ at $t=5.0$. In contrast, EGLM-S and EGLM-C behave similarly; the thermal energy gradually increases, and its distribution is almost uniform. These behaviors of the EGLM schemes are healthy and robust, as the numerically dissipated kinetic and magnetic energies are converted into the thermal energy. On the other hand, CT and GLM can easily produce negative pressure if the plasma beta is even lower. While the stripe patterns look similar between CT and GLM, their detailed evolutions are different, inferring that the main origins of the error are different in CT and GLM. The major part of the error in GLM arises from the ignored energy of the divergence cleaning variable $\psi$. In CT, on the other hand, the cell-centered magnetic fields are calculated by linear interpolation of the face-centered fields, whose error becomes prominent in low-beta regions. This illustrates the weak point of the CT scheme discussed in the Introduction. 

Thus, we recommend including the EGLM source term in the energy equation or using the conservative formula including the energy of $\psi$ explicitly, particularly for simulations including low-beta regions. On the other hand, compared to the energy equation, we believe that the additional EGLM source term in the momentum equation (\ref{eq:eom2}) is of less importance and optional, because this term should remain small as long as the scheme is working well. Moreover, this term violates conservation of the linear momentum as discussed in \citet{dedner}.

\bibliography{ms}{}
\bibliographystyle{aasjournalv7}

\end{document}